\documentclass[
floatfix,
reprint,
superscriptaddress,
amsmath,amssymb,
aps,
prx,
longbibliography]{revtex4-2}

\usepackage{graphicx}
\usepackage{natbib}
\usepackage{amsmath}
\usepackage{amssymb}
\usepackage{appendix}
\usepackage{comment}
\usepackage{relsize}
\usepackage{mathtools}
\usepackage[bottom]{footmisc} 
\usepackage[dvipsnames]{xcolor}
\usepackage[section]{placeins}
\usepackage{layouts}
\usepackage{bbold}
\usepackage{comment}
\definecolor{darkblue}{rgb}{0,0,.65}
\definecolor{darkgreen}{rgb}{0.28,0.41,0.19}

\usepackage[%
    pdfauthor={David J. Luitz},%
  pdfstartview=FitH,%
  breaklinks=true,%
  bookmarks=true,%
  colorlinks=true,%
  anchorcolor=black,%
  citecolor=blue,
  filecolor=black,%
  menucolor=black,%
  urlcolor=darkblue,%
  linkcolor=blue,%
 ]{hyperref}
\usepackage[all]{hypcap} 
\usepackage{braket}

\usepackage{float}

\usepackage{tikz}
\usepackage{tikz-network}

\newcommand{\be}{\begin{equation}}
\newcommand{\ee}{\end{equation}}

\definecolor{Ured}{HTML}{cc0000}
\definecolor{Ublue}{HTML}{1f65cf}
\definecolor{Ugreen}{HTML}{198a11}

\graphicspath{{Plots/}{Plotsnew/}{Fourth_order_quantities/}{tikzfiles/}{FigsJohn/}}
\begin{document}

\title{Eigenstate correlations, the eigenstate thermalization hypothesis, and quantum information dynamics in chaotic many-body quantum systems}

\author{Dominik Hahn}
\email{hahn@pks.mpg.de}
\affiliation{Max Planck Institute for the Physics of Complex Systems, N\"othnitzer Stra{\ss}e~38, 01187-Dresden, Germany}
\author{David J. Luitz}
\email{david.luitz@uni-bonn.de}
\affiliation{Physikalisches Institut, Universit\"at Bonn, Nussallee 12, 53115 Bonn, Germany}
%\affiliation{Max Planck Institute for the Physics of Complex Systems, N\"othnitzer Stra{\ss}e~38, 01187-Dresden, Germany}
\author{J. T. Chalker}
\email{john.chalker@physics.ox.ac.uk}
\affiliation{Theoretical Physics, University of Oxford,
Parks Road, Oxford OX1 3PU, United Kingdom}
\date{\today}%

\begin{abstract}
We consider the statistical properties of eigenstates of the 
time-evolution operator in chaotic many-body quantum systems. Our focus is on correlations between eigenstates that are specific to spatially extended systems and that characterise entanglement dynamics and operator spreading. In order to isolate these aspects of dynamics from those arising as a result of local conservation laws, we consider Floquet systems in which there are no conserved densities. The correlations associated with scrambling of quantum information lie outside the standard framework established by the eigenstate thermalisation hypothesis (ETH). In particular, ETH provides a statistical description of matrix elements of local operators between pairs of eigenstates, whereas the aspects of dynamics we are concerned with arise from correlations amongst sets of four or more eigenstates. We establish the simplest correlation function that captures these correlations and discuss features of its behaviour that are expected to be universal at long distances and low energies. We also propose a maximum-entropy Ansatz
for the joint distribution of a small number $n$ of eigenstates. In the case $n = 2$ this Ansatz reproduces ETH. For
$n = 4$ it captures both the growth with time of entanglement between subsystems, as characterised
by the purity of the time-evolution operator, and also operator spreading, as characterised by the
behaviour of the out-of-time-order correlator. We test these ideas by comparing results from Monte
Carlo sampling of our Ansatz with exact diagonalisation studies of Floquet quantum circuits.
\end{abstract}

\maketitle

%%%%%%%%%%%%%%%%%%%%%%%%%%%
\section{Introduction}
%%%%%%%%%%%%%%%%%%%%%%%%%%%

Although textbook approaches to the thermodynamic equilibrium of quantum systems rely on invoking a weak coupling to a heat bath, 
it was understood over the course of the last four decades that this is not strictly necessary. 
In a large class of many-body quantum systems, the interactions between its constituents enable an isolated system to act as its own heat bath and to reach a thermal equilibrium state at long times when starting from most nonequilibrium initial states. 
While the late-time state remains a pure state, it approaches a \emph{typical state}, representative of the Gibbs ensemble with small fluctuations away from this state for large system size \cite{peres_ergodicity_1984}.
This phenomenon emerges from the pseudo-random nature of physical observables in the energy-eigenbasis, which was suggested as a criterion for quantum chaos \cite{peres_ergodicity_1984} and demonstrated numerically early on \cite{feingold_ergodicity_1984}, with diagonal matrix elements clustering around equilibrium expectation values \cite{feingold_distribution_1986}. Integrable systems can evade this behaviour, but are not robust in the sense that very weak perturbations suffice to recover thermalisation \cite{deutsch_quantum_1991}.

These observations were subsequently formalised and now constitute the eigenstate thermalisation hypothesis (ETH) \cite{peres_ergodicity_1984,feingold_ergodicity_1984,deutsch_quantum_1991,srednicki_chaos_1994,rigol_thermalization_2008,beugeling_finite-size_2014,beugeling_off-diagonal_2015,mondaini_eigenstate_2017,khaymovich_eigenstate_2019,alba_eigenstate_2015,dalessio_quantum_2016,borgonovi_quantum_2016}, which can be derived from the assumption that the behaviour of the quantum system within a narrow energy window is essentially captured by random matrix theory. This generically leads to Gaussian distributions of the matrix elements of local operators in the energy eigenbasis \cite{Prosen1999,beugeling_off-diagonal_2015,mondaini_eigenstate_2017,khaymovich_eigenstate_2019}, with possible exceptions in tails of the distribution for systems with slow particle transport \cite{luitz_long_2016,luitz_anomalous_2016,serbyn_thouless_2017,roy_anomalous_2018}. In these Gaussian distributions, the mean values of diagonal matrix elements are fixed to reproduce statistical mechanical averages of observables, while off-diagonal matrix elements have zero mean and an energy structure in their variance that governs the dynamics of autocorrelation functions of the local operators \cite{khatami_fluctuation-dissipation_2013,dalessio_quantum_2016}.

It was pointed out recently on general grounds that this picture cannot be complete \cite{foini_eigenstate_2019,Chan2019Eigenstate}. If matrix elements of local operators in the eigenbasis were independent Gaussian random variables, then their mean and variance would determine not only autocorrelation functions, but also all higher-order correlators. In particular, the implications for the out-of-time-order correlators (OTOC) \cite{larkin_quasiclassical_1969,shenker_black_2014} are in stark contrast to the known behaviour of chaotic many-body quantum systems. This leads to the conclusion that there are necessarily correlations between matrix elements, 
which contain information characterised by higher order cumulants \cite{foini_eigenstate_2019,Chan2019Eigenstate,brenes_out--time-order_2021,jafferis_matrix_2023}. 

More specifically, in quantum systems with local interactions, there are strong bounds on how fast correlations can spread \cite{lieb_finite_1972}, and this limits for instance the rate of growth of the entanglement entropy \cite{calabrese_evolution_2005,kim_ballistic_2013}. 
%Independently of the choice of the initial state, 
This behaviour is reflected in the typically linear growth of the operator entanglement entropy of the time evolution operator of local systems \cite{zhou_operator_2017,dubail_entanglement_2017}, and is captured by light-cone structures of out-of-time-order correlators \cite{maldacena_bound_2016}. Our focus here is on the resulting correlations between matrix elements of observables, and related correlations between eigenstates of the time-evolution operator.

A separate potential source of correlations between matrix elements is provided by locally conserved densities. Such correlations were considered early on \cite{prosen_statistical_1994}, have been investigated via transport timescales \cite{dymarsky_bound_2022} and subsequently observed in larger systems \cite{richter_eigenstate_2020, wang_eigenstate_2022}.
To isolate features arising because of the dynamics of quantum information from those due to locally conserved densities, we focus in the following on Floquet systems, for which time-dependence of the Hamiltonian eliminates energy conservation, and in which there are no other local conservation laws.

%%%% Free probability
Recent work has presented a generalised formulation of ETH using Free Probability theory and numerical tests for higher-order correlations between matrix elements
%based on a generalized formulation of ETH using free probability theory~
\cite{pappalardi_eigenstate_2022,pappalardi_general_2023}. 
That perspective considers matrix elements of local operators as fundamental objects,
revealing the frequency structure of the higher-order free cumulants, particularly 
%$k_4$,
the fourth-order free cumulant, which encodes the leading
%non-trivial 
correlations of matrix elements beyond standard ETH. 

An alternative perspective, which we adopt here and which was considered for example in Ref.~\cite{shi_local_2023}, is to treat eigenstates of the time-evolution operator, rather than matrix elements of observables, as the relevant set of variables. In particular, the typical time evolution of Renyi entropy in local systems implies non-trivial correlations between eigenstates. 
Separately, in Ref.~\cite{liao_field_2022} a derivation is given of ETH via a study of eigenstates in random Floquet quantum circuits using a field-theoretic approach.

In the present paper, we discuss correlations in chaotic many-body quantum systems that are associated with the dynamics of quantum information scrambling and that can be expected to be universal in spatially extended systems with local interactions. Here we use the term \emph{chaotic} to refer to generic systems whose dynamics is unconstrained by (for example) integrability, many-body localization or Hilbert space fragmentation. While ETH is formulated as a statement about the statistical properties of matrix elements of operators, we find that it is more transparent to consider instead eigenstates and correlators constructed from them, without reference to particular operators.
The first main contribution of our work is to identify the leading-order eigenstate correlator that contains information about this 
scrambling dynamics, and to discuss its behaviour at long distances and low energy differences. The second main contribution is to introduce an Ansatz for the joint probability distribution of a small number $n$ of eigenstates of the time-evolution operator. In the case of a pair of eigenstates, this reproduces the Gaussian distribution for matrix elements of local observables that constitutes ETH. Extending this Ansatz, we show that 
%a quartic term, 
a constraint on the joint distribution of four eigenstates is sufficient to capture the essential features of the OTOC and of the operator entanglement entropy of the time-evolution operator. As a third main contribution, we test these ideas by comparing results from exact diagonalisation of the time-evolution operator with those from Monte Carlo sampling of our Ansatz for the joint eigenstate distribution.

The approach that we develop here builds on previous work by one of the present authors and others \cite{Chan2019Eigenstate}, which argued for the existence of eigenstate correlations on the basis of known behaviour of the OTOC, and demonstrated their presence using numerical studies of Floquet quantum circuits. Our aim in the following is to establish a general framework for the discussion of these phenomena via the three contributions summarised in the preceding paragraph, none of which were anticipated in \cite{Chan2019Eigenstate}.

The ideas and results that we set out in this paper are consistent with, but broadly complementary to, the recent discussion of a generalised version of ETH, formulated to describe the average of products of arbitrary numbers of matrix elements of observables using the language of Free Probability \cite{pappalardi_eigenstate_2022,pappalardi_general_2023}. In particular, our emphasis is different from that of \cite{pappalardi_general_2023} in two ways. First, we focus on correlations at large distances and small energy separations that we expect to be a universal consequence of the dynamics of quantum information. Second, we find that it is advantageous to consider correlations between eigenstates in place of matrix elements. 

The remainder of this paper is organised as follows. In Sec.~\ref{sec:Synopsis} we provide a compact overview of our main results. In Sec.~\ref{sec:Implementation} we give details of our calculations, including the microscopic models we use for numerical studies, the determination of parameters in our Ansatz for the joint probability distribution of eigenstates, and a summary of the numerical methods used.
We develop a treatment of our Ansatz based on a perturbative expansion in Sec.~\ref{sec:perturbation theory} and we present numerical results for additional models in Sec.~\ref{sec:further models}. We conclude with a summary and outlook in Sec.~\ref{sec:Discussion and outlook}.

%%%%%%%%%%%%%%%%%%%%%%%%%%%%
\section{Overview}\label{sec:Synopsis}
%%%%%%%%%%%%%%%%%%%%%%%%%%%%

In this section we provide an overview of our results. We introduce the class of model studied and the eigenstate correlators of interest in Sec.~\ref{subsec2A}. We set out the relationship between these correlators and the OTOC in Sec.~\ref{relntoOTOC}, and indicate generalisations in Sec.~\ref{subsec:multi}. We review the sense in which the original form of ETH fails to capture these correlators in Sec.~\ref{subsec2B} and we propose a representation for them in terms of the joint distribution function for sets of four eigenstates in Sec.~\ref{subsec2B2}. We present results for the correlators from exact diagonalisation and from Monte Carlo sampling of this distribution in Sec.~\ref{subsec2C}.

%%%%%%%%%%%%%%%%%%%%%%%%%%%%%%%%%%%%%%%%%%%%%%%%%%%%%%%%%%%%
\subsection{Models and Correlation Functions}\label{subsec2A}
%%%%%%%%%%%%%%%%%%%%%%%%%%%%%%%%%%%%%%%%%%%%%%%%%%%%%%%%%%%%%

We start by setting out some essential notation. We consider a one-dimensional Floquet system consisting of $L$ sites, each with a local Hilbert space dimension $q$, coupled by local interactions. We use $W$ to denote the Floquet operator for the system, which is then a unitary $q^L\times q^L$ matrix that generates evolution through one time period. Defining eigenstates $\ket{a}$ and quasienergies $\theta_a$ satisfying ${W} |a\rangle = \mathrm{e}^{-\mathrm{i}\theta_a} |a\rangle$, 
the evolution operator $W(t)$ 
for an integer number of time-steps $t$ has the spectral decomposition 
\begin{equation}
W(t)=\sum_a \mathrm{e}^{-\mathrm{i}\theta_a t} |a\rangle\langle a|\,.
\end{equation} 
The dynamics can be characterised in terms of correlators of local operators $X_\alpha, Y_\beta \ldots$. Here we use upper case letters $X, Y, \ldots$ as labels for subsystems on which local operators act, with subscripts $\alpha, \beta \ldots$ to distinguish different operators acting on a given subsystem. Later we will use $\overline{X}$, $\overline{Y}, \ldots$ to denote the complements of these subsystems. The time evolution in the Heisenberg picture is $X_\alpha(t) = W^\dagger(t)X_\alpha W(t)$, and for a Floquet system it is natural to evaluate correlators using the infinite-temperature density matrix.
As we describe in more detail in Sec.~\ref{subsec:Models}, the models analysed in this work consist of local gates drawn from the Haar ensemble.
In this case, it is useful to consider an ensemble of realisations of $W$ and to average physical quantities over the ensemble. We indicate this average by $[\ldots ]_{\rm av}$. An alternative average is over a Haar distribution of vectors, which we indicate by $[\ldots ]_{0}$.

The simplest correlator is the autocorrelation function of a single operator, which has the spectral decomposition 
\begin{equation}\label{autocorrelation}
    q^{-L}{\rm Tr} [X_\alpha(t) X_\alpha] = q^{-L} \sum_{ab} |\langle a|X_\alpha|b\rangle|^2\mathrm{e}^{\mathrm{i}(\theta_a - \theta_b) t}\,.
\end{equation}
In a chaotic many-body quantum system this is expected to decay on a timescale that is microscopic in the sense that it is of order a few Floquet periods~\cite{chan_eigenstate_2019}. Evidently, its behaviour reflects statistical properties of pairs $|a\rangle, |b\rangle$ of eigenstates, which are therefore expected to show features as a function of the quasienergy difference $\theta_a - \theta_b$ that vary on a scale only a few times smaller than the spectral width $2\pi$ \cite{chan_eigenstate_2019}.

%By contrast, for the OTOC
The OTOC has the definition and spectral decomposition
\begin{eqnarray}\label{OTOCdef}
    q^{-L}{\rm Tr}[X_\alpha(t) Y_\beta X_\alpha(t) Y_\beta]=  q^{-L} &&
    \sum_{abcd} \langle a| X_\alpha |b\rangle \langle b| Y_\beta | c \rangle \times \nonumber\\
    \times \langle c | X_\alpha |d\rangle \langle d|Y_\beta |a \rangle && e^{i(\theta_a - \theta_b +\theta_c - \theta_d)t}\,.
\end{eqnarray}
If the subsystems $X$ and $Y$ are separated by a large distance, the main features of the OTOC appear on a large timescale. More specifically, the support of the operator $X_\alpha(t)$ is expected \cite{lieb_finite_1972} to grow with a butterfly velocity $v_{\rm B}$; the OTOC is constant and non-zero if this support is disjoint from that of $Y_\beta$, but falls towards zero when the support of $X_\alpha(t)$ expands to contain that of $Y_\beta$.
%the main features appear on a timescale that is large if $X$ and $Y$ are well-separated subsystems. 
Clearly, behaviour of the OTOC reflects statistical properties of sets of four eigenstates, $|a\rangle, |b\rangle, |c\rangle$ and $|d\rangle$, which must show features on a quasienergy scale $2\pi v_{\rm B}/s$, with $s$ being the spatial separation of the operators, that is much smaller for large $s$ than the spectral width.

Besides the OTOC, a second way to characterise quantum information dynamics is via the spread of entanglement. Consider an initial state $|\psi\rangle$ with low entanglement in the site basis. Although the corresponding density matrix $\rho(t) = W(t)|\psi\rangle \langle \psi|W^\dagger(t)$ remains pure at all $t$, for a subsystem $X$ that is much smaller than its complement $\overline{X}$, the reduced density matrix $\rho_X(t) = {\rm Tr}_{\overline{X}} \rho(t)$ is expected to evolve towards an equilibrium one. This is probed at the simplest level by considering the purity ${\rm Tr}_X [\rho_X(t)]^2$. Since the definition of the purity involves two powers of $W(t)$ and two of $W^\dagger(t)$, its behaviour, like that of the OTOC,  reflects correlations amongst sets of four eigenstates. These are characterised by the correlation function defined below in Eq.~\eqref{F2}. %implies {\color{blue}\sout{non-trivial}} correlations between quadruples of eigenstates, which we introduce in the following.}

Both the OTOC and the purity require choices in their definitions -- of the operators denoted by $X_\alpha$ and $Y_\beta$ in Eq.~(\ref{OTOCdef}) for the former, and of the initial wavefunction $|\psi\rangle$ for the latter. This arbitrariness can be eliminated by averaging the OTOC over two complete sets of operators $\{X_\alpha\}$ and $\{Y_\beta\}$ with given supports $X$ and $Y$, and by considering the operator entanglement entropy of $W(t)$ \cite{prosen_operator_2007,zhou_operator_2017,dubail_entanglement_2017} in place of the purity of $\rho_X(t)$. Both routes lead to an identical correlator which is defined solely in terms of sets of four eigenstates and the choice of $X$ and $Y$. We defer discussion of details and present first an alternative argument that singles out the same correlator.

As a starting point, consider the Schmidt decomposition of an eigenstate in terms of tensor products of orthonormal basis states $|i_X\rangle$ and $|i_{\overline{X}}\rangle$ for subsystem $X$ and its complement $\overline{X}$, which we write as
\begin{equation}\label{Schmidt}
    |a\rangle = \sum_{i_Xi_{\overline{X}}} [C_X(a)]_{i_Xi_{\overline{X}}} |i_X\rangle \otimes |i_{\overline{X}}\rangle\,.
\end{equation}
%\djl{corrected $C_x \to C_X$ above}
Here $C_X(a)$ is the matrix version of the eigenstate $\ket{a}$, separating states on subsystem $X$ into row indices and states on $\overline{X}$ into column indices. It hence has dimensions $q^{L(X)}\times q^{L(\overline{X})}$, where $L(X) = |X|$, the number of sites in $X$, and similarly for $L(\overline{X})$. %\djl{harmonize notation: $L(X)=|X|$ like we use later?}

\begin{figure}[t!]
	\centering
	\includegraphics[width=0.40\textwidth]{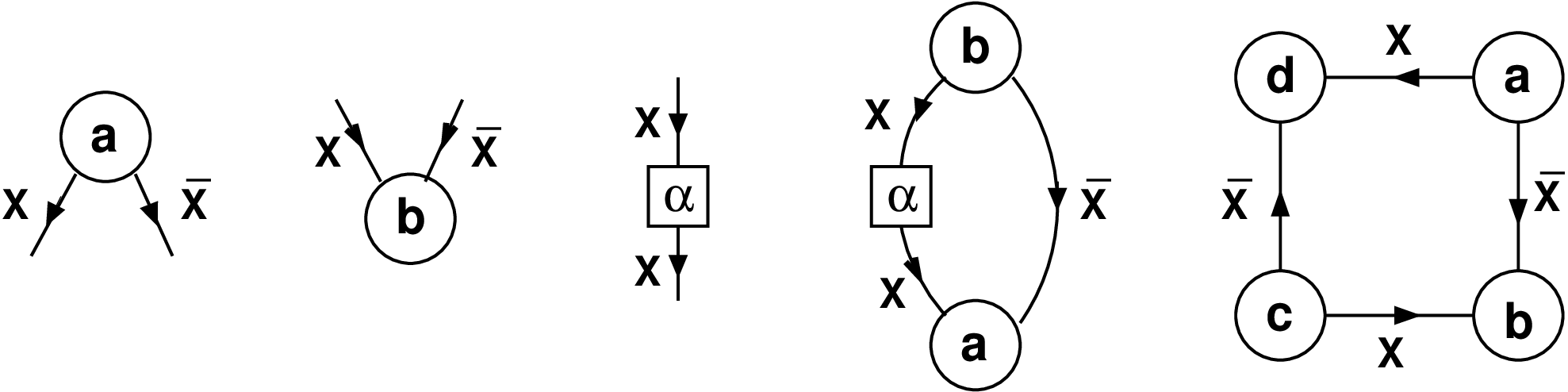}
    \caption{Ingredients for a diagrammatic representation: from left to right $C_X(a)$, $C_X(b)^\dagger$, $X_\alpha$, $\langle a|X_\alpha| b\rangle$, and $M_X(abcd)$.}
	\label{fig:one}
\end{figure}

The problem of constructing correlators from sets of eigenstates is equivalent to one of forming scalars from sets of matrices $C_X(a)$. This can be done by taking the trace of products of an even number of terms, in which the matrices 
%(in general, rectangular) 
alternate with their Hermitian conjugates. The matrices within such a trace must all refer to a given choice of subsystem $X$ but may refer to multiple eigenstates $|a\rangle$, $|b\rangle \ldots$. At lowest order this recipe simply yields the quantity 
\begin{equation}\label{ortho}
{\rm Tr}[C_X(a) C^\dagger_X(b)]=\delta_{ab},
\end{equation}
which has a value fixed by orthonomality of the eigenstates. At next order it gives
\begin{equation}\label{M}
    M_X(abcd) = {\rm Tr}[C_X(a)C^\dagger_X(b) C_X(c) C_X^\dagger(d)]\,.
\end{equation}
Such quantities can be represented diagrammatically as shown in Fig.~\ref{fig:one}

We now invoke two guiding ideas. One follows from the fact that we want to characterise dynamics in space and time, which suggests that we should consider more than one way of subdividing the system into subsystems, with different alternatives labelled $X$, $Y$, $\ldots$. The other follows from the fact that the overall phases of individual eigenstates can be chosen arbitrarily, but physical quantities should be invariant under the transformation $|a\rangle \to e^{i\phi_a}|a\rangle$. To eliminate the phases $\phi_a$, each $C_X(a)$ appearing in a correlator must be accompanied by a Hermitian conjugate $C_Y^\dagger(a)$, referring to the same eigenstate but possibly with a different division into subsystems. Employing both these ideas, we are led to one of the two main quantities of interest in this paper,
$M^{\phantom{*}}_X(abcd)M^*_Y(abcd)$ and its appropriately normalised ensemble average
\begin{eqnarray}\label{F2}
    F_4(X,Y,\theta) =  && q^{-L(X,Y)}\Big[\sum_{abcd}M^{\phantom{*}}_X(abcd)M_Y^*(abcd)\times \nonumber\\
   &&\times \delta(\theta - \theta_a + \theta_b -\theta_c +\theta_d)\Big]_{\rm av}\,.
\end{eqnarray}
Here and in the following, the argument of the $\delta$-function on quasienergy differences is taken modulo $2\pi$. The length $L(X,Y)$ appearing in the normalisation is defined by 
\begin{eqnarray}
    L(X,Y) %&=& 2|X\setminus\overline{Y}| + 2|\overline{X}\setminus{Y}|+|\overline{X}\setminus\overline{Y}|+|X\setminus Y|\nonumber \\
    &=& 2L - |\overline{X}\setminus\overline{Y}| - |X\setminus Y|\nonumber \\
 &=&  2L - |\overline{Y}\setminus\overline{X}| - |Y\setminus X| = L(Y,X)\,,
\end{eqnarray}    
where notation of the form $|A\setminus B|$ indicates the number of sites that are in subsystem $A$ but not in $B$. The choice of subscript on $F_4(X,Y,\theta)$ indicates that this quantity characterises correlations within sets of four eigenstates. The normalization in Eq.~\eqref{F2} is chosen such that the Fourier transform of Eq.~\eqref{F2} with respect to the relative phase $\theta$ is unity at $t=0$, as will be explained below Eq.~\eqref{eq:time}.

Somewhat surprisingly, at this order in powers of the eigenstates, Eq.~\eqref{F2} is the unique outcome  of interest from the approach we have sketched for constructing correlators. To see this, consider potential alternatives. Any such alternative should involve two factors, each consisting of a trace over products of Schmidt matrices $C_X(a)$, since each trace carries a subsystem label and we are interested in correlations between a pair of subsystems. Moreover, at this order the two factors together involve four such matrices and four Hermitian conjugates. If these matrices are equally distributed between the two traces, then alternatives to Eq.~\eqref{F2} must all be generated by replacing $M_Y^*(abcd)$ with a similar factor that preserves invariance under changes of eigenstate phases. These can all be obtained from Eq.~\eqref{F2} using the equalities $M_Y(abcd) = M_Y(cdab) = M_{\overline{Y}}(adcb) = M^*_Y(badc)$ and so are equal to $F_4(X,Y,\theta)$ or $F_4(X,\overline{Y},\theta)$. Finally, one might regroup matrices under the trace, so that one trace involves six matrices and the other trace involves only two. Then, however, the value of the second trace is fixed via Eq.~\eqref{ortho} and is independent of subsystem label, eliminating the spatial dependence of interest.

Amongst correlators involving only a single $X$, the lowest order quantity that is independent of eigenstate phases is $M_X(abba)$ (equal to $M_{\overline{X}}(aabb)$). From this we define
\begin{eqnarray}\label{F1}
    F_2(X,\theta) &=& q^{-(L+L(X))} \times \nonumber\\ &\times& \Big[\sum_{ab} M_X(abba) \delta(\theta - \theta_a + \theta_b) \Big]_{\rm av}\,.
\end{eqnarray}
%where $L(X) = |X|$, the number of sites in subsystem $X$. 
The subscript on $F_2(X,\theta)$ indicates that this correlator characterises correlations between pairs of eigenstates.

The two correlators $F_2(X,\theta)$ and $F_4(X,Y,\theta)$ are the central quantities of interest in the following, together with their counterparts in the time domain, defined by
\begin{equation}
    f_2(X,t) = \int_{-\pi}^{\pi} {\rm d}\theta \, F_2(X,\theta) e^{i\theta t}
\end{equation}
and
\begin{equation}\label{eq:time}
f_4(X,Y,t) =  \int_{-\pi}^{\pi} {\rm d}\theta \, F_4(X,Y,\theta) e^{i\theta t}\,.
\end{equation}

The initial values $f_2(X,0) = f_4(X,Y,0)=1$ follow from completeness of the set of eigenstates and the choices of normalisation in Eqns.~\eqref{F2} and \eqref{F1} (see Eqns.~\eqref{eq:sumrule1} and \eqref{eq:sumrule2} for a discussion). The late-time limits are also system independent. For $f_4(X,Y,t)$ this limit comes (assuming no degeneracies) from terms in \eqref{F2} with pairwise equal labels $a{=}b, c{=}d$ or $a{=}d, b{=}c$. For $f_2(X,t)$ it comes from terms in \eqref{F1} with $a{=}b$. 
All these terms can be simplified by noting that 
$C_X(a)C_X^\dagger(a)\equiv {\rm Tr}_{\overline{X}}|a\rangle\langle a|$ is the reduced density matrix on subsystem $X$ formed from the eigenstate $|a\rangle$. If $L(X) \ll L$, one expects from ETH that to an excellent approximation $C_X(a)C^\dagger_X(a) = q^{-L(X)}\openone_X$, where $\openone_X$ is the identity on $X$. This implies that $\lim_{|t|\to\infty} f_2(X,t) = q^{-2L(X)}$ and that $\lim_{t\to\infty} f_4(X,Y,t) = q^{2L - L(X,Y) -L(X) - L(Y)}$ for $L(X), L(Y) \ll L$. 

%%%%%%%%%%%%%%%%%%%%%%%%%%%%%%%%%%%%%%%%%%%%%%%%%%%%%%%%%%%%%%%%%%%%%%%%%%%%%%%%%%%%%%%%%%%%%%%%%%%%%%%%%%%%%%%%%%
\subsection{Relation to autocorrelation functions of observables and OTOC}\label{relntoOTOC}%%%%%%%%%%%%%
%%%%%%%%%%%%%%%%%%%%%%%%%%%%%%%%%%%%%%%%%%%%%%%%%%%%%%%%%%%%%%%%%%%%%%%%%%%%%%%%%%%%%%%%%%%%%%%%%%%%%%%%%%%%%%%%%%

\begin{figure}[t!]
	\centering
	\includegraphics[width=0.35\textwidth]{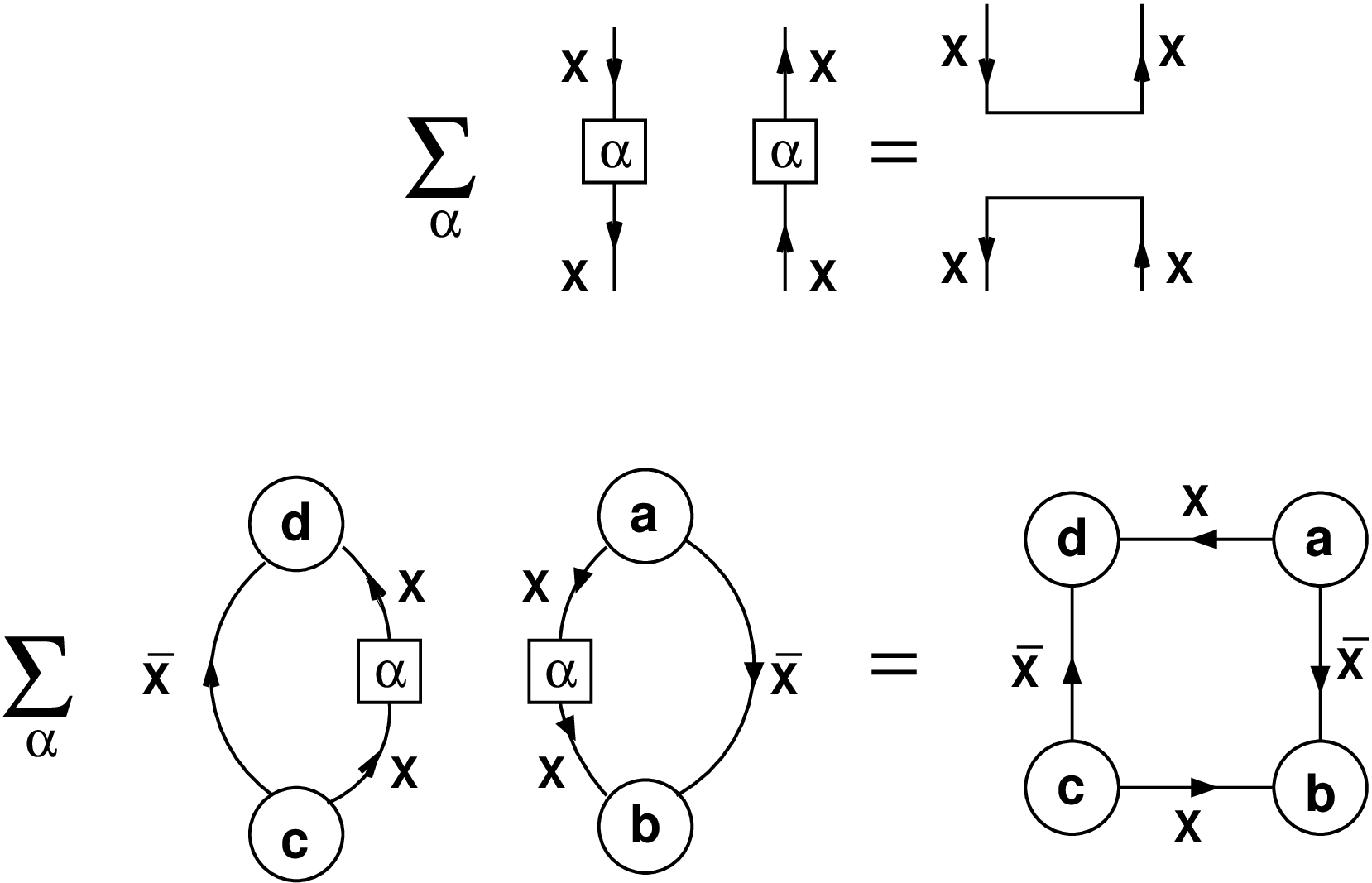}
    \caption{Diagrammatic representation of: (top) the operator resolution of the identity; (bottom) Eq.~\eqref{res}.}
	\label{fig:two}
\end{figure}

As we now discuss, these correlators are related respectively to the autocorrelation function [Eq.~\eqref{autocorrelation}] and the OTOC [Eq.~\eqref{OTOCdef}] via averages over the operators appearing in the latter two quantities. We begin by stating a key relation between $M_X(abcd)$ and an operator average of matrix elements. Given a subsystem $X$, choose a complete basis of $q^{2L(X)}$ Hermitian operators $\{X_\alpha\}$ that act on the subsystem and obey the orthonormality condition 
\begin{equation}\label{norm}
   q^{-L(X)} {\rm Tr}_X [X_\alpha X_\beta] = \delta_{\alpha \beta}\,. 
\end{equation}
Using the resolution of the identity in the vector space of operators, one finds
\begin{equation}\label{res}
    q^{-L(X)} \sum_\alpha \langle a|X_\alpha |b\rangle \langle c| X_\alpha |d\rangle = M_X(abcd)\,.
\end{equation}
The operator resolution of the identity and the relation given in Eq.~\eqref{res} are represented diagrammatically in Fig.~\ref{fig:two}.

We first apply this to the simpler case of the autocorrelation function. The autocorrelation function averaged over all choices of operator (and over the ensemble of systems) is 
\begin{equation}\label{eq:F2relation}
    q^{-2L(X)}\sum_\alpha \Big[q^{-L} {\rm Tr}[X_\alpha(t) X_\alpha] \Big]_{\rm av} = f_2(X,t)\,.
\end{equation}
The special case $X_\alpha = \openone_X$ is the only contribution to this average that survives at late times, giving a value of $\lim_{|t|\to\infty}f_2(X,t)$ consistent with the discussion above.

Similar arguments apply to the OTOC. In this case we choose two complete sets of operators. Operators in one set act on the subsystem labelled $X$ and are denoted by $X_\alpha$. Operators in the other set act on the subsystem labelled $Y$ and are denoted by $Y_\beta$. Then
%, using the resolution of the identity in the vector spaces spanned by each of the sets of operators, one finds 
\begin{eqnarray}
    q^{-(L(X)+L(Y))}&&\sum_{\alpha \beta} \langle a| X_\alpha |b\rangle \langle b| Y_\beta | c \rangle 
     \langle c | X_\alpha |d\rangle \langle d|Y_\beta |a \rangle \nonumber \\
     &&= M_X(abcd) M_Y(bcda)\,.
\end{eqnarray}
This can be rewritten using $M_Y(bcda) = M^*_{\overline{Y}}(abcd)$, and so the average of the OTOC over both sets of operators is
\begin{eqnarray}\label{eq:connectionOtoc}
    q^{-2(L(X)+L(Y))}&&\sum_{\alpha \beta} \Big[ q^{-L}{\rm Tr}[X_\alpha(t) Y_\beta X_\alpha(t) Y_\beta]\Big]_{\rm av} \nonumber\\ &&= q^{-S(X,Y)} f_4(X,\overline{Y},t)
\end{eqnarray}
with $S(X,Y) = L(X) + L(Y) +L - L(X,\overline{Y}) = |X \setminus \overline{Y}| + |Y \setminus\overline{X}|$. Contributions to this average from the special cases $X_\alpha = \openone_X$ and/or $Y_\beta = \openone_Y$ survive at long times and are responsible for the limiting value given above.

The correlator $f_4(X,Y,t)$ also arises from a discussion of the operator entanglement entropy of the evolution operator. This quantity stems from considering the operator $W(t)$ as a state on a doubled Hilbert space, with components given by the matrix elements $[W(t)]_{i_Xi_{\overline{X}},j_Yj_{\overline{Y}}}$. The corresponding reduced operator density matrix, obtained by tracing out the degrees of freedom in the subsystems $\overline{X}$ and $\overline{Y}$, is
\begin{eqnarray}
[\rho(X,Y,W(t))&&]_{i_Xj_Y,l_Xm_Y}= \sum_{ab} [C_X(a)C_X^\dagger(b)]_{i_Xl_X} \times \nonumber\\
\times &&[C_Y(b)C_Y^\dagger(a)]_{m_Yj_Y} e^{i(\theta_b - \theta_a)t}\,.
\end{eqnarray}
The ensemble-averaged operator purity arising from this reduced density matrix is simply
\begin{equation}\label{operatorpurity}
    \big[{\rm Tr} [\rho(X,Y,W(t))^2]\big]_{\rm av} = q^{L(X,Y)}f_4(X,Y,t)\,.
\end{equation}

The proportionality between $f_4(X,Y,t)$ and the operator purity of the evolution operator implies a straightforward link to the idea of an entanglement membrane, which has been proposed as a coarse-grained description of entanglement dynamics in chaotic many-body quantum systems \cite{Nahum2017Quantum,Jonay2018Coarse-grained}. For the one-dimensional models we are considering, the entanglement membrane is a curve in space-time, and to discuss the link to operator purity we build on the exposition of Ref.~\cite{Jonay2018Coarse-grained}. In outline, coarse-grained features of entanglement dynamics are determined by the line tension ${\cal E}(v)$ of this membrane, which is a function of a velocity $v$. In our notation, $v=s/t$, where $s$ is the distance between the ends of subsystems $X$ and $Y$, defined in Fig.~\ref{fig:synopsis}(d). For a fixed choice of $X$ and $Y$ with $s$ large, the operator purity of the evolution operator is proportional to the line tension, and so traces out the function ${\cal E}(v)$ as $t$ varies. Hence the correlators $f_4(X,Y,t)$ and $F_4(X,Y,\theta)$ can be seen as representations of the line tension ${\cal E}(v)$. 

%%%%%%%%%%%%%%%%%%%%%%%%%%%%%%%%%%%%%%%%%%%%%%%%%%%%%%%%%%%%%%%%%%%%%%%%%%%%%%%%
\subsection{Multi-time and multi-quasienergy correlators}\label{subsec:multi}
%%%%%%%%%%%%%%%%%%%%%%%%%%%%%%%%%%%%%%%%%%%%%%%%%%%%%%%%%%%%%%%%%%%%%%%%%%%%%%%%

An obvious generalisation \cite{chan_eigenstate_2019} of the OTOC [Eq.~\eqref{OTOCdef}] is to introduce three time arguments, by considering the quantity $q^{-L}{\rm Tr}[X_\alpha(t+t_2) Y_\beta(t_1) X_\alpha(t) Y_\beta]$. Correspondingly, in the quasienergy domain we have a generalisation of the correlator $F_4(X,Y,\theta)$, defined by
\begin{eqnarray}\label{F2-3}
    F_4(X,Y;\theta,\theta_1,\theta_2) &=& q^{-L(X,Y)}\Big[\sum_{abcd}M^{\phantom{*}}_X(abcd)M_Y^*(abcd)\times \nonumber\\
   &\times& \delta(\theta - \theta_a + \theta_b -\theta_c +\theta_d)\times \nonumber\\
   &\times&\delta(\theta_2 - \theta_a + \theta_b) \times \delta(\theta_1 - \theta_b + \theta_c)\Big]_{\rm av}
\end{eqnarray}
with the Fourier transform
\begin{eqnarray}
f_4(X,Y;t,t_1,t_2) &=& \int_{-\pi}^\pi \!\!\!\!\!{\rm d}\theta \int_{-\pi}^\pi \!\!\!\!\!{\rm d}\theta_1 \int_{-\pi}^\pi \!\!\!\!\!{\rm d}\theta_2\,\, F_4(X,Y;\theta,\theta_1,\theta_2)\times \nonumber\\ &\times& e^{i(\theta t + \theta_1 t_1 + \theta_2 t_2)}
\end{eqnarray}
which is related to the generalised OTOC by
\begin{eqnarray}
f_4(X,\overline{Y};t,t_1,t_2) &=& q^{S(X,Y)-2L(X) - 2L(Y)} \times \nonumber \\ \times \sum_{\alpha \beta}q^{-L}&{\rm Tr}&[X_\alpha(t+t_2) Y_\beta(t_1) X_\alpha(t) Y_\beta]\,.
\end{eqnarray}
The single-quasienergy correlator can be recovered from the multi-quasienergy version using
\begin{equation}
F_4(X,Y,\theta) = \int_{-\pi}^\pi {\rm d}\theta_1 \int_{-\pi}^\pi {\rm d}\theta_2 \,\,F_4(X,Y;\theta,\theta_1,\theta_2)\,.
\end{equation}
\begin{comment}
We also have 
\begin{equation}
F_2(X,\theta_2) = \int_{-\pi}^\pi {\rm d}\theta \int_{-\pi}^\pi {\rm d}\theta_1 \,\,F_4(X,Y;\theta,\theta_1,\theta_2)
\end{equation}
and 
\begin{equation}
F_2(Y,\theta_1) = \int_{-\pi}^\pi {\rm d}\theta \int_{-\pi}^\pi {\rm d}\theta_2 \,\,F_4(X,Y;\theta,\theta_1,\theta_2)
\end{equation}
\end{comment}

For large separation $s$ between the ends of $X$ and $\overline{Y}$ these correlation functions factorise as follows (see Ref.~\cite{chan_eigenstate_2019}). 
%Chan, De Luca and Chalker,  PRL~{\bf 122}, 220601 (2019)]. 
Consider behaviour in the time domain. Let $v_{\rm B}$ be the butterfly velocity and $D$ the diffusion constant for the spread of an operator front. Then we expect
\begin{eqnarray}
%f_4(X,\overline{Y};t,t_1,t_2)
 &q^{-L}&{\rm Tr}[X_\alpha(t+t_2) \overline{Y}_\beta(t_1)X_\alpha(t) \overline{Y}_\beta]\nonumber\\  
 &\approx& \left\{\begin{array}{lll} 
 q^{-L}{\rm Tr}[X_\alpha(t+t_2) X_\alpha(t) ]\times &&\\ \quad \times q^{-L}{\rm Tr}[ \overline{Y}_\beta(t_1)  \overline{Y}_\beta]
%f_2(X,t_2) f_2(Y,t_1) 
& \,\, & s - v_{\rm B} |t| \gg \sqrt{Dt}\\
0 & & v_{\rm B} |t| - s \gg \sqrt{Dt} \nonumber
\end{array}\right.
\end{eqnarray}
This motivates the approximation
\begin{equation}
f_4(X,{Y};t,t_1,t_2) \approxeq f_4(X,{Y},t) f_2(X,t_2) f_2({Y},t_1)\,,
\end{equation}
which in the quasienergy domain is
\begin{equation}\label{eq:multi}
F_4(X,{Y};\theta,\theta_1,\theta_2) \approxeq F_4(X,{Y},\theta) F_2(X,\theta_2) F_2({Y},\theta_1)\,.
\end{equation}
For the models we study in this paper, $F_2(X,\theta_2)$ is only weakly dependent on $\theta$ and so the multi-quasienergy correlator carries only limited extra information compared to the single-quasienergy version. For this reason we leave study of $F_4(X,{Y};\theta,\theta_1,\theta_2)$ for future work.

%%%%%%%%%%%%%%%%%%%%%%%%%%%%%%%%%%%%%%%%%%%%%%%%%%%%%%%%%%%%%%%%%%%%%%%%%%%%%%
 \subsection{Existence of correlations beyond ETH}\label{subsec2B}
%%%%%%%%%%%%%%%%%%%%%%%%%%%%%%%%%%%%%%%%%%%%%%%%%%%%%%%%%%%%%%%%%%%%%%%%%%%%%%

Our objective in the remainder of this work is to find a form for the joint distribution function (JDF) of a small number of eigenstates that reproduces these correlations. We do this using a maximum entropy Ansatz with a final form that we build up by considering first individual vectors, then pairs of vectors, and finally sets of four vectors.

To place our approach in context, it is useful to recall (following Refs.~\cite{foini_eigenstate_2019} and \cite{chan_eigenstate_2019}) the limitations of ETH in its standard formulation when applied to the OTOC. As a starting point, consider the spectral decomposition of the OTOC in terms of operator matrix elements, as displayed in Eq.~\eqref{OTOCdef}. ETH asserts that matrix elements of the form $\langle a_1|X_\alpha|a_2\rangle$ and $\langle a_3|Y_\beta|a_4 \rangle$ appearing in this expression are Gaussian random variables, and are independent apart from the constraint implied by Hermiticity of the operators $X_\alpha$ and $Y_\beta$. The mean values of off-diagonal matrix elements are automatically zero, and those of diagonal matrix elements are zero for traceless operators in the Floquet setting of interest. Finally, the variance of these matrix elements is set by the Hilbert space size and is ${\cal O}(q^{-L})$. 

Applying these ideas to Eq.~\eqref{OTOCdef}, the OTOC is given by $q^{-L}$ times a sum of $q^{4L}$ random ${\cal O}(q^{-2L})$ terms. Of these, only the $q^L$ terms with $a{=}b{=}c{=}d$ are expected from ETH to have a non-zero average. This would imply an average value for the OTOC of ${\cal O}(q^{-2L})$. Treating the remaining terms as independent random variables, one expects ${\cal O}(q^{-L})$ fluctuations around this average. In contrast, the true value is ${\cal O}(1)$ at short times. To resolve this discrepancy it is necessary that a product of four matrix elements of the form $\langle a|X_\alpha|b\rangle \langle b| Y_\beta| c \rangle \langle c| X_\alpha | d\rangle \langle d|Y_\beta |a\rangle$ has a non-zero average that is ${\cal O}(q^{-3L})$ in addition to the ${\cal O}(q^{-2L})$ fluctuations captured by the standard version of ETH \cite{chan_eigenstate_2019}. These additional correlations are the central concern in this paper and in the generalisation of ETH discussed in Refs.~\cite{pappalardi_eigenstate_2022,pappalardi_general_2023}. 

A simple demonstration that such correlations must be present, regardless of details of the dynamics, is provided by a sum rule related to the value of the OTOC at $t=0$. From the left-hand side of Eq.~\eqref{OTOCdef}, assuming for simplicity that the subsystems $X$ and $Y$ do not overlap and using the operator normalisation of Eq.~\eqref{norm}, we have
\begin{equation}\label{eq:sumrule1}
    q^{-L}{\rm Tr}[X_\alpha(t)Y_\beta X_\alpha(t) Y_\beta] \Big|_{t=0} = 1\,.
\end{equation}
Using this in Eq.~\eqref{eq:connectionOtoc} with Eq.~\eqref{eq:time} we have
\begin{equation}\label{eq:sumrule2}
  \int_{-\pi}^{\pi} {\rm d}\theta \, F_4(X,\overline{Y},\theta) = 1\,.
\end{equation}
This sum rule for $F_4(X,\overline{Y},\theta)$ is automatically satisfied if eigenstates are Haar distributed vectors, and in that case $F_4(X,\overline{Y},\theta)$ is independent of $\theta$. The eigenstate correlations that we are concerned with generate a dependence of $F_4(X,\overline{Y},\theta)$ on $\theta$ but do not alter the fact that, with the normalisation of Eq.~\eqref{F2}, it has an order of magnitude that is independent of the Hilbert space dimension $q^L$. 

%%%%%%%%%%%%%%%%%%%%%%%%%%%%%%%%%%%%%%%%%%%%%%%%%%%%%%%%%%%%%%
 \subsection{Describing correlations beyond ETH}\label{subsec2B2}
%%%%%%%%%%%%%%%%%%%%%%%%%%%%%%%%%%%%%%%%%%%%%%%%%%%%%%%%%%%%%%

 Some constraints on the eigenstate JDF are implied by ETH, which we now consider.
 ETH specifies statistical properties of both diagonal and off-diagonal matrix elements of local observables between eigenstates, and we discuss the two classes of matrix elements separately. 

For a system with a time-independent Hamiltonian, a key part of ETH is that diagonal matrix elements of observables vary smoothly with energy, taking average values compatible with a thermal ensemble at the same energy density, and with fluctuations of a characteristic size that vanishes rapidly as the thermodynamic limit is approached. By contrast, for Floquet systems, statistical properties of diagonal matrix elements of observables are independent of quasienergy.
We capture this property of diagonal matrix elements in a Floquet system by taking individual eigenstates to be isotropically distributed vectors in the Hilbert space for the model. 
We outline in Sec.~\ref{sec:Discussion and outlook} the alternative choice required to model the energy dependence of diagonal matrix elements in Hamiltonian systems within our approach. 
We denote the isotropic (Haar) distribution for one, two or four orthonormal vectors by $P_1^{(0)}(a)$, $P_2^{(0)}(a,b)$ and $P_4^{(0)}(a,b,c,d)$ respectively, and set out to modify these distributions in a way that introduces the correlations of interest. 

Statistical properties of off-diagonal matrix elements determine the approach to equilibrium and the autocorrelation functions of observables. ETH applied to Floquet systems asserts that these matrix elements are independent Gaussian random variables with a variance that depends only on quasienergy separation. A central idea in our work is that strict independence is incompatible with the correlations implied by the dynamics of quantum information. Instead there are correlations (albeit weak) and these are better handled by considering distributions for eigenstates rather than matrix elements. We take the joint distribution of a pair of eigenstates to have the Maximum Entropy form 
\begin{equation}\label{P2}
    P_2(a_1,a_2) = Z_2^{-1} P_2^{(0)}(a_1,a_2) e^{ - S_2(a_1,a_2)}
\end{equation}
with $Z_2$ a normalisation constant and
\begin{eqnarray}\label{LagrangeG2}
    S_2(a,b) = \sum_X G_2(X,\theta_a - \theta_b) M_X(abba)\,,
\end{eqnarray}
where the coefficients $G_2(X,\theta)$ act as a Lagrange multipliers and should be chosen to reproduce the behaviour of $F_2(X,\theta)$ as determined for a particular system. 

Extending this pattern, we take the joint distribution of four eigenstates to have the form
\begin{eqnarray}\label{JDF}
    P_4(a_1,a_2,a_3,a_4) &&= Z_4^{-1} P_4^{(0)}(a_1,a_2,a_3,a_4) \times \nonumber\\ 
    &&\times e^{ - \sum_{j<k} S_2(a_j,a_k)-S_4(a_1,a_2,a_3,a_4)}
\end{eqnarray}
with $Z_4$ a normalisation constant and
\begin{eqnarray}\label{LagrangeG4}
    S_4(a,b,c,d) &=& \sum_{XY} G_4(X,Y,\theta_a-\theta_b+\theta_c-\theta_d)\times \nonumber\\ &&\times M_X(abcd) M^*_Y(abcd)\,.
\end{eqnarray}
Here the Lagrange multipliers $G_4(X,Y,\theta)$ should be chosen to reproduce the behaviour of $F_4(X,Y,\theta)$. 

Two further ingredients are required. One is to establish a practical method for deducing the values of the Lagrange multipliers from information on the correlators. The other is to test the approach by sampling $P_2(a,b)$ or $P_4(a,b,c,d)$ and comparing the results with correlators calculated from exact diagonalisation (ED) of $W$. In this work we focus on the correlator $F_4(X,Y,\theta)$ since it contains the long-distance, low-energy information related to the dynamics of quantum information. Moreover, in the models we study, the lower-order correlator $f_2(X,t)$ decays rapidly in time. This implies that $F_2(X,\theta)$ is approximately independent of quasi-energy and so we simply set $G_2(X,\theta)$ to zero in our initial treatment. We return to consideration of a $\theta$-dependent $F_2(X,\theta)$ and non-zero $G_2(X,\theta)$ immediately after our discussion of $F_4(X,Y,\theta)$.
%in Sec.~\ref{sec:F_2}.

The determination of the Lagrange multipliers from ED data for eigenstate correlator can be seen simply as a fitting problem, but since this involves a high-dimensional parameter space, alternative approaches are desirable.  Fortunately, as we describe in Sec.~\ref{sec:probfunction}, we have been able to find direct and straightforward methods to derive $G_4(X,Y,\theta)$ from $F_4(X,Y,\theta)$ and $G_2(X,Y,\theta)$ from $F_2(X,Y,\theta)$.

%%%%%%%%%%%%%%%%%%%%%%%%%%%%%%%%%%%%%%%%%%%%%%%%%%%%%%%%%%%%%%%%%%%
\subsection{Results}\label{subsec2C}
%%%%%%%%%%%%%%%%%%%%%%%%%%%%%%%%%%%%%%%%%%%%%%%%%%%%%%%%%%%%%%%%%%%

\begin{figure*}[t!]
    \begin{center}\hspace{1cm}
  \includegraphics[width=0.96\textwidth]{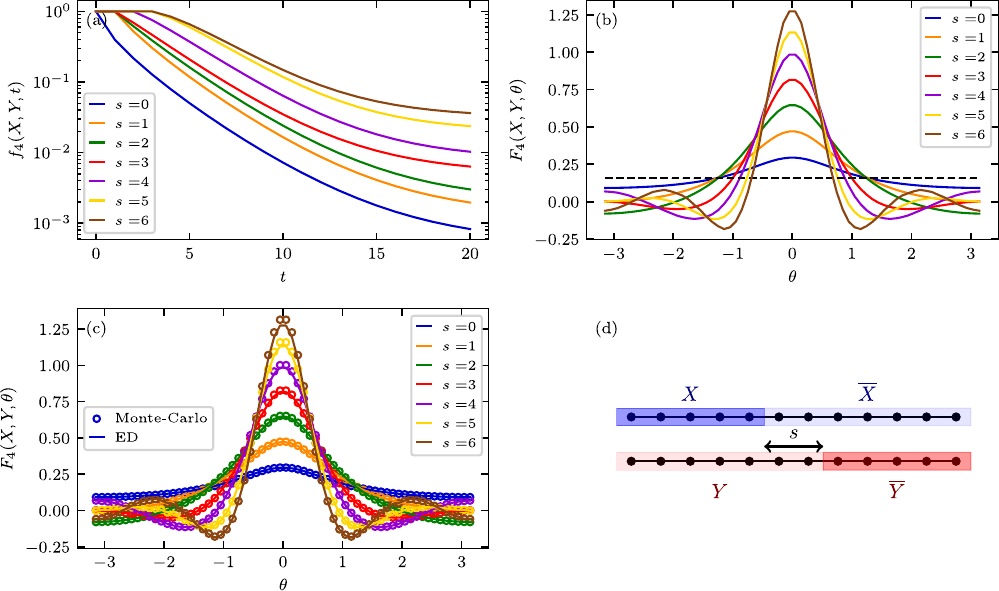}
\end{center}
    \caption{Overview of main results, calculated for the brickwork circuit model defined in Sec.~\ref{subsec:Models} with $L=12$, $q=2$ and open boundary conditions: behaviour from ED in (a) and (b); comparison of ED and MC in (c); and geometries of subsystems $X$ and $Y$ in (d).
    (a) The correlator $f_4(X,Y,t)$ [Eq.~\eqref{F1}] vs $t$, 
    %for $X$ and $Y$ as illustrated in (d), 
    obtained using ED. 
    Recall [Eq.~\eqref{eq:connectionOtoc}] that this correlator is proportional to the OTOC $q^{-L}{\rm Tr}[X_\alpha(t) \overline{Y}_\beta X_\alpha(t) \overline{Y}_\beta]$ averaged over operators $X_\alpha$ and $\overline{Y}_{\beta}$ with support on subsystems $X$ and $\overline{Y}$ respectively. Decay of the correlator reflects operator spreading, and the onset time for decay increases with $s$. 
    (b) The correlator $F_4(X,Y,\theta)$ [Fourier transform of $f_4(X,Y,t)$: see Eq.~\eqref{F2}] vs quasienergy difference $\theta$, obtained using ED [contributions to Eq.~\eqref{F2} in which any of the state labels $a$, $b$, $c$ and $d$ are equal have been omitted: they are atypical and carry vanishing weight in the thermodynamic limit]. It has a peak centred on $\theta=0$ which grows narrower and higher with increasing $s$, reflecting the short-time plateau in (a); black dashed line: behaviour when the Floquet operator $W$ is modelled using a $q^L\times q^L$ Haar unitary, showing for this structureless case that $F_4(X,Y,\theta)$ is non-zero but $\theta$-independent.  
    (c) Comparison of ED results with MC results from the Ansatz for the JDF of Eq.~\eqref{JDF} fitted to behaviour in the geometries of (d), showing excellent agreement between $F^{\rm MC}_4(X,Y,\theta)$ (open circles from MC) and $F_4(X,Y,\theta)$ (lines from ED) vs $\theta$ for various $s$. %with $X$ and $Y$ as in (d).
    (d) Illustration of two ways of dividing the 12-site system with open boundary conditions into subsystems by means of a single spatial cut. In one case the subsystems are labelled $X$ and $\overline{X}$; in the other the labels are $Y$ and $\overline{Y}$. The distance between the spatial cuts in the two cases is denoted by $s$. 
    }
	\label{fig:synopsis}
\end{figure*}

We implement and test these ideas using an open chain with the Floquet operator defined by a brickwork circuit (see Sec.~\ref{subsec:Models}). 
 If non-zero Lagrange multipliers $G_4(X,Y,\theta)$ were included for all choices of subsystem $X$ and $Y$, their number would be unreasonably large, both as a matter of principle and in practice (there are $2^L$ different subsystems in total). Since our models have nearest-neighbour interactions, it seems natural to organise subsystems according to the number of cuts necessary to separate them from the full system. For the majority of our work we include non-zero Lagrange multipliers only for subsystems that can be obtained from a full system under open boundary conditions by means of a single spatial cut. We discuss alternative choices of the sets of subsystems $X$ and $Y$ for which the Lagrange multipliers $G_4(X,Y,\theta)$ are non-zero in Sec.~\ref{sec:probfunction} and Appendix \ref{sec:doublecut}.  As an additional restriction on the choice of non-zero Lagrange multipliers, our fastest method (see Sec.~\ref{G4fit}) for determining their values is effective if the probability distribution of $M_X(abcd)$ is well-approximated by a Gaussian, which is the case for the model studied provided the subsystem sizes $L(X)$ and $L(\overline{X})$ are not too small. In order to satisfy this requirement, and in order to limit the total number of Lagrange multipliers under consideration, we include $G_4(X,Y,\theta)$ for all $L-5$ subsystems $X$ with $L(X)>2$ that can be obtained from the full system by means of a single spatial cut, and similarly for $Y$. Taking account of symmetry under interchange of $X$ and $Y$, this gives $(L-5)(L-4)/2$ (i.e. 28 for the example studied of $L=12$) independent Lagrange multipliers.
Using as input the values of $F_4(X,Y,\theta)$ for these subsystem choices from ED, we arrive at a final form for the JDF. 

We use Monte Carlo (MC) sampling of this distribution to compute the correlator [denoted by $F^{\rm MC}_4(X,Y,\theta)$] with two objectives. First, for the simplest choices of $X$ involving single spatial cuts (and similarly for $Y$), comparison with ED is a test of our Ansatz for the JDF and of our procedure to determine the Lagrange multipliers. Second, it is interesting to see whether this input alone is sufficient to capture correlations more generally. To probe this we compare ED and MC results for the correlator, making choices of $X$ (and also $Y$) that are defined by more than one spatial cut. This is a test of the extent to which the proposed JDF captures long-distance, low energy correlations in general. In particular, taking $X$ and $Y$ each to consist of a small number of sites acting as the support for a local observable, we test the implications of the JDF for the (operator-averaged) OTOC. 

\begin{figure*}[t!]
 \centering
  \includegraphics[width=0.9\textwidth]{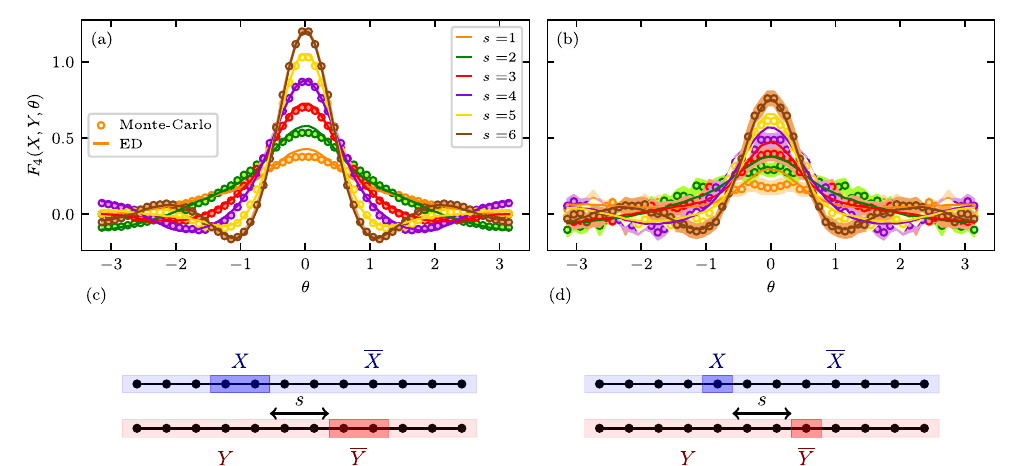}
    \caption{Test of JDF fitted to behaviour in the geometries of Fig.~\ref{fig:synopsis}(d) but applied to geometries of Fig.~\ref{fig:synopsisOtoc}(c) and Fig.~\ref{fig:synopsisOtoc}(d), respectively. (a) and (b) Comparison of data from MC (open circles) and ED (solid lines). 
    (c) Partition used for (a), in which the 12-site system is divided by two spatial cuts into a two-site subsystem $X$ and its complement $\overline{X}$, or a two-site subsystem $\overline{Y}$ and its complement $Y$. 
    (d) Partition used for (b), in which the 12-site system is divided by two spatial cuts into a single-site subsystem $X$ and its complement $\overline{X}$, or a single-site subsystem $\overline{Y}$ and its complement $Y$.
    Calculations are for the brickwork circuit model defined in Sec.~\ref{subsec:Models} with $L=12$, $q=2$ and open boundary conditions.
    }
	\label{fig:synopsisOtoc}
\end{figure*}

Some of our principal results are shown in Fig.~\ref{fig:synopsis} and discussed in the figure caption. The main conclusions are as follows. (i) As expected from its relation to the OTOC, the correlator $f_4(X,Y,t)$ is time-independent at short times and falls off at a timescale that is long if the spatial separation $s$ between subsystems $X$ and $\overline{Y}$ is large. All aspects of this behaviour are apparent in Fig.~\ref{fig:synopsis}(a). (ii) In turn, this implies structure in $F_4(X,Y,\theta)$ at small quasienergies $\theta$, as is visible in Fig.~\ref{fig:synopsis}(b); the width in quasienergy of this structure decreases with increasing $s$. (iii) Monte Carlo sampling of the JDF of Eq.~\eqref{JDF}, with Lagrange multipliers determined as described in Sec.~\ref{sec:probfunction}, generates results for $F_4(X,Y,\theta)$ that are in excellent agreement with those from ED, as demonstrated in Fig.~\ref{fig:synopsis}(c).

Further important results are shown in Fig.~\ref{fig:synopsisOtoc}. Here we examine how well the JDF constructed using correlators for single-cut subsystems can capture correlators for two-cut subsystems. It is apparent from Fig.~\ref{fig:synopsisOtoc}(a) that MC sampling of the JDF generates a moderately good representation of $F_4(X,Y,\theta)$ for the two-site choices of $X$ and $Y$ shown in Fig.~\ref{fig:synopsisOtoc}(c). Equivalently, the JDF determined using information from the geometries of Fig.~\ref{fig:synopsis}(d) reproduces the main features of the OTOC, as a function of time and spatial separation, for two operators, each supported on two sites in the geometry of Fig.~\ref{fig:synopsisOtoc}(c). A similar picture holds even when the support of the operators is reduced to a single site, as is shown in Fig.~\ref{fig:synopsisOtoc}(b). In this case, statistical errors are enhanced, since in each subsystem there are only one quarter as many operators to average over.

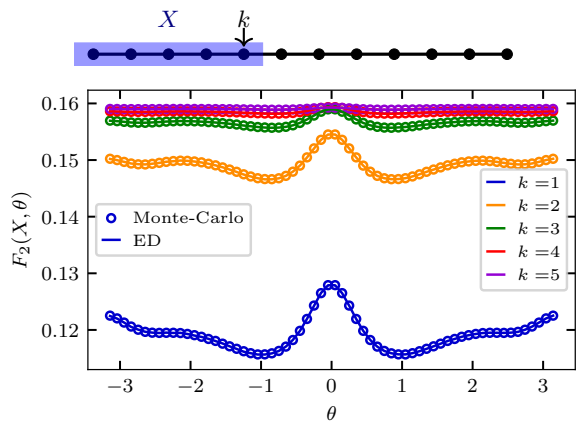
\begin{figure}[h]
	\centering
    %\documentclass{standalone}
%\usepackage{tikz}
%\begin{document}
        \begin{tikzpicture}
        \draw[very thick] (0.5,0) -- (6,0);
        \foreach \x in {0.5,1.0,1.5,2.0,2.5,3.0,3.5,4.0,4.5,5.0,5.5,6.0}
            \filldraw (\x,0) circle (2pt);
            \filldraw[opacity=0.4,color=blue] (0.25,-0.15) rectangle (2.75,0.15);
            %\filldraw[opacity=0.4,color=red] (3.75,-0.15) rectangle (6.25,0.15);
            %\draw[very thick,<->] (2.75,0.25) -- node[above] {$s$} (3.75,0.25);
            \node[anchor=south,color=blue!50!black] (x) at (1.5,0.25) {$X$};
\node[anchor=south,color=blue!0!black] (k) at (2.5,0.25) {$k$};
\draw[thick][->] (2.5,0.35) to     (2.5,0.1);
            
            %\node[anchor=south,color=red!50!black] (y) at (5,0.25) {$Y$};
    \end{tikzpicture}
%\end{document}        
    \vspace{0.4cm}
	\includegraphics[width=0.45\textwidth]{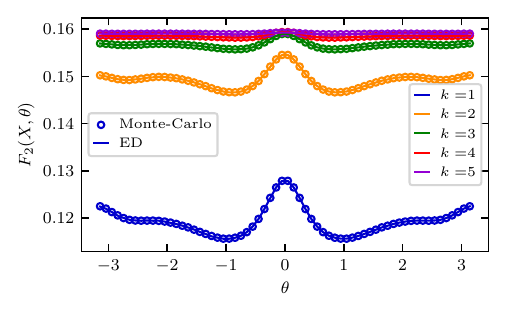}
    \caption{ Numerical results for $F_2(X,\theta)$ in a system with $L=12$ and $q=2$ as a function of the left cut position $k$: ED results~(solid lines) vs. Monte-Carlo~(circles).  [Delta-function contributions at $\theta=0$ have been omitted; they are responsible for ensuring that the sum rule $\int_{-\pi}^{\pi} {\rm d}\theta\, F_2(X,\theta) = 1$ is satisfied for all $X$.]  The almost perfect agreement between both sets of data indicates that the correlations between pairs of eigenstates $F_2(X,\theta)$, as captured by ETH, may also be represented accurately within our approach.}
	\label{fig:S2-secII}
\end{figure}

As a final indication of the effectiveness of our approach, we return to the behaviour of $F_2(X,\theta)$, which characterises the correlations that are incorporated by ETH. For simplicity we consider only the joint distribution of pair of eigenstates, in this way treating $F_2(X,\theta)$ separately from $F_4(X,Y,\theta)$. By determining the Lagrange multiplier $G_2(X,\theta)$ from ED data as described in Sec.~\ref{sec:probfunction} we generate and sample from this joint distribution, with results that are shown in Fig.~\ref{fig:S2-secII}. As is evident, our MC data are in excellent agreement with ED results for all values of quasienergy difference $\theta$ and all subsystem choices $X$. 

The fact that deviations are small from the value $F_2(X,\theta) = (2\pi)^{-1}\approxeq 0.159$ for a pair of Haar-distributed orthogonal unit vectors is justification for our omission of $F_2(X,\theta)$ in our discussion of $F_4(X,Y,\theta)$. A more complete treatment would require the simultaneous inclusion of both $G_2(X,\theta)$ and $G_4(X,Y,\theta)$, following Eq.~\eqref{JDF}.
Some consequences have been discussed previously in Ref.~\cite{chan_eigenstate_2019} and we do not consider them further here. 

The overall aim of MC sampling from our Maximum Entropy Ansatz for the JDF of a small number of eigenstates is to test whether, with a suitable choice of Lagrange multipliers, the JDF reproduces the correlations $F_4(X,Y,\theta)$ measured from ED. This test of our approach is a crucial one, and we believe Fig.~\ref{fig:synopsis}(c) offers excellent evidence that the JDF can indeed reproduce the required correlations for the model and parameter range investigated there. Further discussion of the determination of Lagrange multipliers, including a treatment of other models, is presented in Sec.~\ref{sec:further models} and the Appendix.

\section{Models, Lagrange multipliers and numerical methods}\label{sec:Implementation}

 In Sec.~\ref{subsec:Models} we describe the microscopic models we use to generate the numerical results shown in this paper. In Sec.~\ref{sec:probfunction} we set out efficient methods to determine the Lagrange multipliers $G_2(X,\theta)$ and $G_4(X,Y,\theta)$ that appear in the JDF for eigenstates. In Sec.~\ref{sec:Numerical details} we give details of the numerical methods used in this paper.
\subsection{Models}
\label{subsec:Models}
\begin{figure}[t]
    \begin{center}
%    U_{(i_0 i_1 \dots), (j_0 j_1,\dots)}=
    \definecolor{Ured}{HTML}{cc0000}
\definecolor{Ublue}{HTML}{1f65cf}
\definecolor{Ugreen}{HTML}{198a11}
%\begin{document}
\begin{tikzpicture}[baseline={([yshift=-.5ex]current  bounding  box.center)}, scale=1.5]
        {
            \foreach \xx in {-3,...,1}
            {
                {
                    \draw[thick, fill=Ured, rounded corners=2pt] (\xx-0.125,0) rectangle (\xx+0.5+0.125,0+0.25);
                }
                \draw[very thick] (\xx,0) -- (\xx,-0.25);
                \draw[very thick] (\xx+0.5,0) -- (\xx+0.5,-0.25);

                \draw[very thick] (\xx+0.5,0.25) -- (\xx+0.5,0.5);
                \draw[very thick] (\xx,0.25) -- (\xx,0.5);

                \draw[very thick] (\xx+0.5,0.75) -- (\xx+0.5,1);
                \draw[very thick] (\xx,0.75) -- (\xx,1);
            }

            \foreach \xx in {0,2,...,8}
            {
                \node at (\xx/2 -3 -0.125 + 0.75/2, 0.125) {\small $w_{\xx}$};
            }

            %\foreach \xx in {0,...,9}
            %{
            %    \node[anchor=north] at (\xx/2-3,-0.25) {\small $\sigma_{\xx}$};
            %    \node[anchor=south] at (\xx/2-3,1) {\small $\sigma_{\xx}'$};
            %}

            \foreach \xx in {-3,...,0.5}
            {
                \draw[thick, fill=Ublue, rounded corners=2pt] (\xx+0.5-0.125,0.5) rectangle (\xx+0.5+0.5+0.125,0.75);
            }

            \foreach \xx in {1,3,...,9}
            {
                \node[white] at (\xx/2 -3 -0.125 + 0.75/2, 0.5+ 0.125) {\small$w_{\xx}$};
            }

            \draw[very thick] (-3,0.25) -- (-3,1);
            \draw[very thick] (1.5,0.25) -- (1.5,1);
        }
    
\end{tikzpicture}
%\end{document} 
\end{center}
    \caption{Unitary time evolution operator $W$ of the Floquet circuit written as a tensor network. 
    }
    \label{fig:circuit}
\end{figure}
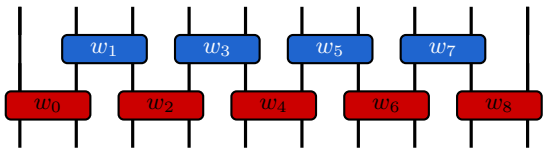

We use the brick-wall circuit depicted in Fig.~\ref{fig:circuit} as a simple model of a periodically driven many-body quantum system with local interactions in one dimension. Such Floquet models have been studied extensively in past work: see for example \cite{Chan2018Solution,chan_eigenstate_2019,Garratt2021Local}. 
This circuit is defined in terms of two-site gates $w_{i}\in \mathbb{C}^{q^2\times q^2}$ and the driving period is decomposed into two parts: in the first half of the period couplings are active only on even bonds of the system, while in the second half the couplings are active only on odd bonds. 
Thus, the time evolution operator for the first half period is  
$W_1 = w_0 \otimes w_2 \otimes w_4 \dots$ and for the second half period is $W_2=\openone\otimes w_1 \otimes w_3 \otimes \dots$.
%, where $\times$ denotes the Kronecker product. 
The evolution operator over one full period is $W = W_2 W_1$, and for $t$ periods we write $W(t)\equiv W^t$.

In order to define an ensemble of systems, a natural choice would be to draw each unitary matrix $w_i$ independently from the Haar distribution. We find, however, that in this case determination of the Lagrange multipliers $G_4(X,Y,\theta)$ is complicated by effects that we attribute to realisations containing weak links $i$ on which the gate $w_i$ is close to the identity (especially in small systems or with small subsystems). To avoid such weak links, we draw the $w_i$ from a truncated version of the Haar distribution in which all gates with an operator purity above a cutoff are discarded. With operator purity defined as in Eq~\eqref{operatorpurity} (so that the two-site identity operator has a purity of $q^4$) we take the cutoff to be $0.3 \times q^4$ for local Hilbert space dimension $q=2$. The consequences of changing or omitting this cutoff are discussed in Sec.~\ref{sec:q2Haarresults} and Appendix~\ref{subsec:Weak links}, and results for $q=3$ are given in Sec.~\ref{sec:q3results}. 

%%%%%%%%%%%%%%%%%%%%%%%%%%%%%%%%%%%%%%%%%%%%%%%%%%%%%%%%%%%%%%%%%%%%%
\subsection{Determining Lagrange multipliers in the JDF}\label{sec:probfunction}
%%%%%%%%%%%%%%%%%%%%%%%%%%%%%%%%%%%%%%%%%%%%%%%%%%%%%%%%%%%%%%%%%%%%%%

We now discuss the problem of determining the Lagrange multipliers that appear in Eq.~\eqref{LagrangeG2} and Eq.~\eqref{LagrangeG4}, and that define the JDFs, Eq.~\eqref{P2} and Eq.~\eqref{JDF}, of a small number of eigenstates. We present a method specific to $G_4(X,Y,\theta)$ in Sec.~\ref{G4fit} and one specific to $G_2(X,\theta)$ in Sec.~\ref{G2fit}. These are both single-shot methods that make explicit use of information about the probability distributions of $M_X(abcd)$ and $M_X(abba)$ respectively, and for this reason they are particularly efficient. In Sec.~\ref{iter} we outline a third, more generally applicable iterative approach, that is agnostic to the probability distributions involved.

%%%%%%%%%%%%%%%%%%%%%%%%%%%%%%%%%%%%%%%%%%%%%%%%
\subsubsection{Determination of $G_4(X,Y,\theta)$}\label{G4fit}
%%%%%%%%%%%%%%%%%%%%%%%%%%%%%%%%%%%%%%%%%%%%%%%%

As noted above, pairwise correlations between eigenstates encoding the decay of autocorrelation functions are very weak in the Floquet model we consider, in the sense that the correlator $F_2(X,\theta)$ is only weakly dependent on $\theta$, taking values close to that for a Haar-distributed pair of orthogonal vectors. As a simplifying approximation, we therefore set to zero the Lagrange multipliers $G_2(X,\theta)$ [see Eq.~\eqref{LagrangeG2}] that control these pairwise correlations.  We make further choices concerning the set of subsystems $X$ and $Y$ for which Lagrange multipliers $G_4(X,Y,\theta)$ are included in the JDF. Without restrictions there are $2^L-1$ distinct subsystems: to reduce this number we include only Lagrange multipliers for connected subsystems -- those that can be obtained from the full system (which has open boundary conditions) by means of a single cut -- and we omit them for all subsystems obtained using multiple cuts.

This choice gives a minimal set of partitions, which turns out to be sufficient to reproduce the main correlations arising from spatial structure and local interactions. As is shown in Fig.~\ref{fig:synopsisOtoc}, it also captures the eigenstate correlations $F_4(X,Y,\theta)$ even for some subsystems pairs $X$ and $Y$ not included in the set. The reason is probably that the most important parameter characterizing the dynamics of information spreading is simply the distance $s$ separating the closest points on the two partitions, and that further possible structure in the partitions is unimportant.

A more general choice is to include Lagrange multipliers for subsystems that can be obtained from the full system by means of either one or two cuts, and we examine this in Appendix~\ref{sec:doublecut}. We find that adding more partitions increases the complexity of calculations without significantly improving the accuracy of MC results for $F_4(X,Y,\theta)$.
Systems with periodic rather than open boundary conditions present a new problem, since in this case a minimum of two cuts is required to define a subsystem. We treat this instance in Sec.\ref{sec:q2resultspbc}.

Our objective is to find the values of the Lagrange multipliers 
$G_4(X,Y,\theta)$ for which the JDF reproduces the eigenstate correlator 
$F_4(X,Y,\theta)$ (known from ED) as accurately as possible. We are able to simplify this task and avoid attacking directly a high-dimensional fitting problem if the quantities $M_X(abcd)$ [Eq.~\eqref{M}] are Gaussian distributed.  
Our motivation for treating a model with a truncated Haar distribution of gate unitaries is that in this system $M_X(abcd)$ is well-approximated by a Gaussian.

\begin{figure}[t!]
	\centering
	\includegraphics[width=0.45\textwidth]{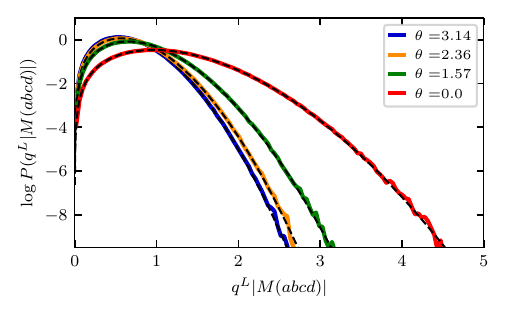}
    \caption{Probability distribution of $q^L |M^{\phantom{*}}_X(abcd)|$ in the Floquet model of Sec.~\ref{subsec:Models} (coloured data) compared with fitted Gaussian distributions (black dashed lines). Data are for $L=12$, $q=2$ and a subsystem $X$ consisting of the four sites closest to the end of an open system, and are shown for the four indicated values of the relative phase $\theta=  \theta_a - \theta_b +\theta_c -\theta_d$. Results for other $q$ and $X$ are shown in Appendix~\ref{sec:Finitesizeeffects}. }
	\label{fig:M0histo}
\end{figure}

Evidence for this is presented in Fig.~\ref{fig:M0histo}. Here we consider only the magnitude $|M_X(abcd)|$ since the phase of $M_X(abcd)$ is dependent on the phase convention used for the eigenstates, and we compare the distribution of $|M_X(abcd)|$ with one in which the real and imaginary parts of $M_X(abcd)$ are assumed to be uncorrelated Gaussian variables with equal variances and zero means. 
An analysis of the dependence of these distributions on the system size $L$ and subsystem size $L(X)$ (see Appendix~\ref{sec:Finitesizeeffects}) suggests that deviations from a Gaussian vanish in the limit of large $L$ and $L(X)$. 
Since deviations are significant for small subsystem size, we omit Lagrange multipliers $G_4(X,Y,\theta)$ for subsystems $X$ with $L(X)$ or $L(\overline{X}) \leq 2$. This leaves $(L-5)^2$ Lagrange multipliers, or $(L-5)(L-4)/2$ independent quantities after taking account of symmetry relations.

Similarly, a Gaussian distribution for $M_X(abcd)$ also arises from the Haar distribution for four orthogonal vectors [$P_4^{(0)}(a,b,c,d)$ in the notation of \eqref{JDF}] in the large $q^L$, $q^{L(X)}$ limit. In this case, and in contrast to the Floquet model, the covariance is independent of the quasienergy difference $\theta$. In the limit of large $q^L$ and $q^{L(X)}$ it has the simple form
\begin{equation}\label{HaarCovariance}
    [M_X(abcd)M^*_Y(abcd)]_{0} \simeq q^{L(X,Y) - 4L}\,.
\end{equation} 
A full expression for Eq.~\eqref{HaarCovariance}, applicable for general $q^L$ and $q^{L(X)}$, can also be obtained in terms of Weingarten functions, but turns out to be unnecessary for the work in this paper, except as noted in Sec.~\ref{sec:q2resultspbc}. 
%\imk{It is possible to obtain an exact analytical expression of Eq.~\eqref{HaarCovariance} using Weingarten functions~\cite{Brouwer1996Diagrammatic}. However, higher-order corrections are negligible for all pairs $X$, $Y$ of partitions we consider and will be thus neglected unless stated otherwise. Furthermore, it turns out that the leading-order result agrees with averaging over a Gaussian ensemble, which will be discussed in  Sec.~\ref{sec:perturbation theory}.}
%{\color{blue}We should discuss this. I have not been able to find a definition of the average $[\ldots ]_0$ in the current version (perhaps I missed it), but my intention was that it should be the Haar average (we have $[\ldots]_{\rm G}$ reserved for the Gaussian averages).}{\color{green}IS the current version sufficient? I do not think that it is possible to find a nice closed form for the next-leading correction, since the order depended on the pair of partitions $X$, $Y$ we choose}

In order to represent these Gaussian distributions in a compact way, it is convenient to introduce notation in which the $L-1$ single-cut subsystems $X$ used to define our Lagrange multipliers are labels for basis states in a $(L-1)$-dimensional vector space. Then the values of $M_X(abcd)$ for different $X$ are components of an $(L-1)$-component column vector $\mathsf{M}$, so that $[\mathsf{M}]_X = M_X(abcd)$, while the Lagrange multipliers are entries in the $(L-1)\times(L-1)$ matrix $\mathsf{G_4}$ with rows and columns labelled by $X$ and $Y$ respectively, so that $[\mathsf{G_4}]_{X,Y}= G_4(X,Y,\theta)$. Similarly, the eigenstate correlators $F_4(X,Y,\theta)$ are elements of a matrix $\mathsf{F_4}$. Then Eq.~\eqref{LagrangeG4} can be rewritten in the compact form
\begin{equation}\label{eq:matrix version}
    S_4 = \mathsf{M}^{\rm T} \mathsf{G_4} \mathsf{M}^*\,.
\end{equation}

The fact that the distributions $P_4(a,b,c,d)$ and $P^{(0)}_4(a,b,c,d)$ are [at large $L$ and $L(X)$] Gaussian for $\mathsf{M}$ suggests that a convenient coordinate system consists of the components of $\mathsf{M}$ together with additional variables $\Omega$ that we do not specify explicitly. We indicate this by writing $P_4({\mathsf{M}},\Omega)$ and  $P^{(0)}_4({\mathsf{M}},\Omega)$. The result given in Eq.~\eqref{HaarCovariance} for the covariance within the Haar distribution implies that
\begin{eqnarray}\label{eq:new distribution function}
    \int {\rm d} \Omega  P^{(0)}_4({\mathsf{M}},\Omega) &=& [{\cal Z}_4^{(0)}]^{-1} e^{-\mathsf{M}^{\rm T} \mathsf{G_4^{(0)} }\mathsf{M}^*}  \\
    \mbox{with} \quad [(\mathsf{G_4^{(0)}})^{-1}]_{X,Y}&=& \big[[\mathsf{M}]_X [\mathsf{M}^*]_Y \big]_{\rm 0} \simeq q^{L(X,Y) -4L}\,.\nonumber
\end{eqnarray}
and ${\cal Z}_4^{(0)} = \pi^{L-1}/ \det{\mathsf{G_4^{(0)}}}$. This in turn implies that
\begin{equation} \label{eq:boundary4}
    \int {\rm d} \Omega P_4({\mathsf{M}},\Omega) = [{\cal Z}_4]^{-1} e^{-\mathsf{M}^{\rm T}[ \mathsf{G_4}+\mathsf{G_4^{(0)}] }\mathsf{M}^*}
\end{equation}
and hence that
\begin{equation} \label{eq:g}
    \big[[\mathsf{M}]_X [\mathsf{M}^*]_Y\big]_{\rm av} =\left[(\mathsf{G_4}+\mathsf{G^0_4})^{-1}\right]_{X,Y}.
\end{equation}
In addition, we have from Eq.~\ref{F2}
\begin{equation}
    [\mathsf{F_4}]_{X,Y} = (2\pi)^{-1}{q^{4L-L(X,Y)}} \big[[\mathsf{M}]_X [\mathsf{M}^*]_Y\big]_{\rm av}\,.
\end{equation}

Eq.~\eqref{eq:g} allows the determination of the Lagrange multipliers in Eq.~\eqref{JDF} in terms of the matrix $\mathsf{F_4}$, which is obtained using ED. This method was used to generate Fig.~\ref{fig:synopsis}(c).

%%%%%%%%%%%%%%%%%%%%%%%%%%%%%%%%%%%%%%%%%%%%%%%%
\subsubsection{Determination of $G_2(X,\theta)$}\label{G2fit}
%%%%%%%%%%%%%%%%%%%%%%%%%%%%%%%%%%%%%%%%%%%%%%%%

Next we describe the method we use to determine the Lagrange multipliers $G_2(X,\theta)$, treating explicitly the case of $L-1$ subsystems $X$ generated by single cuts. A different approach is required to that for $G_4(X,Y,\theta)$ because the probability distribution of $M_X(abba)$ is quite different to that of $M_X(abcd)$ for $a\not=b\not=c\not=d$. Indeed, while (as discussed) $M_X(abcd)$ has a complex Gaussian distribution with zero mean, $M_X(abba)$ is from its definition real  and non-negative. 

To understand the distribution of $M_X(abba)$ it is useful to start from Eq.~\eqref{res}, which specialises here to 
\begin{equation}\label{eq:partitions}
    M_X(abba) =q^{-L(X)} \sum_\alpha |\langle a|X_\alpha |b\rangle|^2\,.
\end{equation}
With $a\not= b$ we expect from ETH that $\langle a|X_\alpha |b\rangle$ for each $\alpha$ is an independent complex Gaussian random variable. From this we can conclude, first, that the two quantities $M_X(abba)$ and $M_{X^\prime}(abba)$ are correlated if the sets $\{X_\alpha\}$ and $\{X^\prime_\alpha\}$ have operators in common, and second, that statistically independent quantities can be constructed using a transformation that organises the operators into disjoint sets.

In detail, this transformation is defined recursively by considering the $L-1$ subsystems in order of increasing size. Again we introduce an $(L-1)$-dimensional vector space, and define the vector $\mathsf{M}$ to have components $[\mathsf{M}]_X = M_X(abba)$. Similarly, we introduce the notation $\mathsf{T}$ for our target vector with statistically independent components $[\mathsf{T}]_X$, and $\mathsf{G}$ with $[\mathsf{G}]_X = G_2(X,\theta)$ for the vector of Lagrange multipliers.  To write the transformation we abuse notation and substitute in place of the component label $X$ the value $\ell=L(X)$. Then using Eq.~\eqref{eq:partitions} we write 
\begin{align}\label{eq:T_i}
\begin{split}
\mathsf{T}_1&=\mathsf{M}_1\\
\mbox{and} \quad \mathsf{T}_{\ell} &=\mathsf{M}_{\ell}-q^{-1} \mathsf{M}_{\ell-1} 
\end{split}
\end{align}
for $\ell=2$ to $L-1$.
This can be recast in the matrix form
\begin{equation}\label{eq:transformationTM}
    \mathsf{T}=\mathsf{V} \mathsf{M}
\end{equation}
where
\begin{equation}\label{eq:transformation}
    \mathsf{V} = \begin{pmatrix} 
    1 & 0 & 0 &\dots &0\\
    -q^{-1}&1&0&\dots&0\\
    \vdots & &\ddots & \\
    0 & 0& \dots   & -q^{-1} & 1 
    \end{pmatrix}\,.
\end{equation}
The effect of this transformation is that Eq.~\eqref{eq:partitions} is replaced by
%\begin{equation}
  $  \mathsf{T}_\ell =q^{-\ell} \sum^\prime_\alpha |\langle a|X_\alpha |b\rangle|^2\,,$
%\end{equation}
where the sum %$\sum^\prime_\alpha$ 
runs over the subset of $n_\ell \equiv (q^2-1)q^{2(\ell - 1)}$ operators $X_\alpha$ that act non-trivially at the rightmost site in $X$ [see illustration in Fig.~\ref{fig:synopsis}(d)].
Since $\langle a|X_\alpha |b\rangle$ is Gaussian with mean zero, the variable $s_\alpha\equiv |\langle a|X_\alpha |b\rangle|^2$ has the distribution $p_\alpha(s_\alpha) = \sigma_\alpha e^{-\sigma_\alpha s_\alpha}$, where $\sigma_\alpha \equiv \big[|\langle a|X_\alpha |b\rangle|^2]_{\rm av}\big]^{-1}$. We make the approximation that $\sigma_\alpha$ takes the same value $\sigma_\ell$ for all $X_\alpha$ that contribute to a given $\mathsf{T}_\ell$. This framework applies not only to the true eigenstate distribution under consideration, but also to vectors with a Haar distribution, and in the latter case we denote the value of $\sigma_\alpha$ by $\sigma^{(0)}_\ell$. Then
\begin{equation}\label{variance}
    [\mathsf{T}_\ell]_{\rm av} = \frac{n_\ell}{\sigma_\ell}
    \quad \mbox{and} \quad 
    [\mathsf{T}_\ell]_{\rm 0} = \frac{n_\ell}{\sigma^{(0)}_\ell}\,.
\end{equation}

With these ingredients in hand, Eq.~\eqref{LagrangeG2} can be written in the form
\begin{equation}\label{S2T}
    S_2 = \mathsf{G}^{\rm T}\mathsf{V}^{-1}\mathsf{T} \quad \mbox{so that} \quad 
    \sigma_\ell = \sigma^{(0)}_\ell + [\mathsf{G}^{\rm T} \mathsf{V}^{-1}]_\ell\,.
\end{equation}
Substituting Eq.~\eqref{variance} into Eq.~\eqref{S2T} and rearranging, we obtain
\begin{equation}\label{G}
    \mathsf{G}_\ell = \sum_{\ell^\prime}\left\{ \frac{n_{\ell^\prime}}{[(\mathsf{V}\mathsf{M})_{\ell^\prime}]_{\rm av}} - \frac{n_{\ell^\prime}}{[(\mathsf{V}\mathsf{M})_{\ell^\prime}]_{\rm 0}}
    \right\}\mathsf{V}_{\ell^\prime \ell}\,.
\end{equation}
We employ Eq.~\eqref{G} to determine the Lagrange multiplier $G_2(X,\theta)$ in terms of $[\mathsf{M}]_{\rm av}$ obtained from ED and the Haar average
\begin{equation}\label{M0}
[\mathsf{M}_\ell]_{\rm 0} = q^{-L}(q^\ell - q^{-\ell})\,. 
\end{equation}
This approach is used to obtain the data shown in Fig.~\ref{fig:S2-secII}.

 Note [from the definition of $\mathsf{T}_\ell$ below Eq.~\eqref{eq:transformation} and the discussion following Eq.~\eqref{variance}] that the probability distribution of $\mathsf{T}_\ell$ is consistent with a Gaussian distribution for the matrix elements $\langle a|X_\alpha| b\rangle$, as expected from ETH, with a variance controlled by the Lagrange multipliers $G_2(X,\theta)$.

%%%%%%%%%%%%%%%%%%%%%%%%%%%%%%%%%%%%%%%%%%%%%%%%
\subsubsection{Iterative method for determining Lagrange multipliers}\label{iter}
%%%%%%%%%%%%%%%%%%%%%%%%%%%%%%%%%%%%%%%%%%%%%%%%

We next describe a straightforward method for determining the Lagrange multipliers without making use of information about the probability distributions of $M_X(abcd)$ and $M_X(abba)$. We treat the case of $G_4(X,Y,\theta)$; the necessary modifications for $G_2(X,\theta)$ are obvious. 

Our starting point is a perturbative expansion of Eq.~\eqref{JDF} to first order in $G(X,Y,\theta)$, which yields
\begin{align}
\begin{split}
    [M_X M_Y^*]_{\rm av}=&[M_X M_Y^*]_{0}\\ &-\sum_{X^\prime Y^\prime}[\mathsf{K}_4]_{XY,X^\prime Y^\prime} G(X^\prime,Y^\prime,\theta)\,,
\end{split}
\end{align}
where $[\mathsf{K}_4]_{XY,X^\prime Y^\prime}$ is the connected correlator
\begin{align}
    [\mathsf{K}_4]_{XY,X^\prime Y^\prime}&=[M_X M_Y^* M_{X^\prime} M_{Y^\prime}^*]_0 \nonumber \\ &\quad -[M_X M_Y^*]_0 [M_{X^\prime} M_{Y^\prime}^*]_0 \,. 
\end{align}
Introducing the abbreviations
\begin{align}
    \begin{split}
        [\mathsf{V}_4]_{XY}&=[M_X M_Y^*]_{0}-[M_X M_Y^*]_{\rm av} \\
      \mbox{and} \quad  [\mathsf{G}_4]_{XY}&=G_4(X,Y,\theta),
    \end{split}
\end{align}
this gives at first order in perturbation theory 
\begin{equation}\label{eq:iterdefiningeq}
    \mathsf{G}_4=\mathsf{K}_4^{-1}\mathsf{V}_4.
\end{equation}

To go beyond first order perturbation theory, we define an iterative procedure based on Eq.~\eqref{eq:iterdefiningeq}. Let $[M_XM^*_Y]^{(n)}_{\rm MC}$ denote the MC result obtained with the $n$th approximant $G_4^{(n)}(X,Y,\theta)$ as Lagrange multiplier, and let $[M_XM^*_Y]_{\rm ED}$ be the value from ED. Then iterate for $n=1, 2 \ldots$
\begin{align}
[\mathsf{V}_4^{(n+1)}]_{XY} = [M_XM^*_Y]^{(n)}_{\rm MC} - [M_XM^*_Y]_{\rm ED} 
\end{align}
with 
\begin{align}
[\mathsf{V}_4^{(1)}]_{XY} = [M_XM^*_Y]_0 - [M_XM^*_Y]_{\rm ED} \,.
\end{align}
and
\begin{align}\label{eq:iterate}
\mathsf{G}^{(n)}_4 = \mathsf{G}^{(n-1)}_4 + \mathsf{K}_4^{-1} \mathsf{V}^{(n)}_4\,.
\end{align}
with
\begin{align}
\mathsf{G}^{(0)}_4 =0\,.
\end{align}

A possible refinement is to replace Eq.~\eqref{eq:iterate} with
\begin{align}\label{eq:alpha}
\mathsf{G}^{(n)}_4 = \mathsf{G}^{(n-1)}_4 + \alpha \mathsf{K}_4^{-1} \mathsf{V}^{(n)}_4\,.
\end{align}
where $0< \alpha \leq 1$ is a real parameter. Small $\alpha$ reduces the risk of overshooting the solution at the expense of a slower convergence rate.

This method is used to produce the data shown in Fig.~\ref{fig:CorrelationsL8q2Haar}.

%%%%%%%%%%%%%%%%%%%%%%%%%%%%%%%%%%%%%%%%%%%%%%%%%%%%%%%%%%%%%%%%%%%
\subsection{Numerical methods}\label{sec:Numerical details}%%%%%%%%
%%%%%%%%%%%%%%%%%%%%%%%%%%%%%%%%%%%%%%%%%%%%%%%%%%%%%%%%%%%%%%%%%%%

In this section, we give details of the ED, MC and ensemble averaging procedures used to obtain the data presented in this paper. 

%%%%%%%%%%%%%%%%%%%%%%%%%%%%%%%%%%%%%%%%%%%%%%%%%%%%%%%%%%%%%%%%%
\subsubsection{Exact diagonalization}\label{sec:ED}%%%%%%%%%%%%%%
%%%%%%%%%%%%%%%%%%%%%%%%%%%%%%%%%%%%%%%%%%%%%%%%%%%%%%%%%%%%%%%%%

 In order to study eigenstate correlations, we use ED of the Floquet operator $W$ to compute exact eigenvectors and hence averages of $M^{\phantom{*}}_X(abcd)M_Y^*(abcd)$ for all choices of our selected subsystems $X$ and $Y$. We obtain phase-resolved averages by dividing the phase interval $[-\pi,\pi]$ into $64$ bins.
 Each realisation of $W$ with $L$ spins generates approximately $q^{4L}/4!$ different quadruples of eigenvectors and from these we randomly choose up to $10^6$ quadruples.
 We average over between 150 (for $L=12$) and 1000 (for $L=10$ and $L=8$) realisations of $W$ to obtain the ED data in Fig.~\ref{fig:synopsis},
 %and Fig.~\ref{fig:M0rel}, 
 and over 25000 realisations for the results shown in Fig.~\ref{fig:synopsisOtoc}. The symmetry relation $F_4(X,Y,\theta)=F_4(X,Y,-\theta)$ allows us to restrict calculations to $\theta\geq 0$.
 Similarly, in the case of $M_X(abba)$ we take $10^6$ tuples out of the approximately $q^{2L}/2$ possibilities for each realisation of $W$ and $L=12$. We average over 1000 realisations for $L=12$ to obtain the data in Fig.~\ref{fig:S2-secII}. 

 From these ED results we compute the Lagrange multipliers $G_4(X,Y,\theta)$ and $G_2(X,\theta)$ using the procedures described in Sec.~\ref{sec:probfunction}. We find that a further symmetrisation of the data, using the spatial symmetry under the interchange of $X$, $Y$ with $\overline{X}$, $\overline{Y}$, improves stability.

As discussed in section~\ref{subsec2B}, the coherent contribution of the four-point correlator $\langle a| X_\alpha |b\rangle \langle b| Y_\beta | c \rangle
    \times \langle c | X_\alpha |d\rangle \langle d|Y_\beta |a \rangle$ is suppressed by a factor $q^{-L}$ in comparison to fluctuations of this quantity. To average out the latter, it is thus necessary to average over $\mathcal{O}(q^{2L})$ samples. In practice, averaging over the operators $X_\alpha$ and $Y_\beta$ reduces fluctuations, but for large spatial separation $s$ and system size $L$, the number of required samples scales exponentially with the system size for a fixed support of $X$ and $\overline{Y}$.
%%%%%%%%%%%%%%%%%%%%%%%%%%%%%%%%%%%%%%%%%%%%%%%%%%%%%%%%%%%%%%%%%%%%%
\subsubsection{Monte-Carlo sampling}\label{sec:Monte-Carlo}%%%%%%%%%%
%%%%%%%%%%%%%%%%%%%%%%%%%%%%%%%%%%%%%%%%%%%%%%%%%%%%%%%%%%%%%%%%%%%%%

In order to determine the eigenstate correlator $F_4(X,Y,\theta)$ from the JDF [Eq.~\eqref{JDF}] for four eigenstates $|a\rangle$, $|b\rangle$, $|c\rangle$ and $|d\rangle$  we use Monte Carlo sampling with $e^{-S_4(a,b,c,d)}$ [Eq.~\eqref{LagrangeG4}] as the weighting term. Similarly, to determine the correlator $F_2(X,\theta)$ from the JDF [Eq.~\eqref{P2}] for two eigenstates $|a\rangle$ and $|b\rangle$  we use Monte Carlo sampling with $e^{-S_2(a,b)}$ [Eq.~\eqref{LagrangeG2}] as the weighting term. 

%In the first instance 
We follow the Metropolis-Hastings algorithm to obtain a Markov chain of vector quadruples $(a,b,c,d)$ distributed according to $P_4$. To generate the next quadruple $(a',b',c',d')$, we randomly select one of the four vectors $a,b,c,d$ and add an vector $\sqrt{\epsilon} v$ to the chosen state, where $v$ is a Haar random vector orthogonal to $a,b,c,d$.

We set $\epsilon=0.1$ for $L=12$ and $q=2$, and $\epsilon=0.8$ for $L=8$ and $q=2$. Since $V$ is unitary, orthonormal vectors retain this property after the transformation.

This choice of update rule with tuning parameter $\epsilon$ allows us to perform effective importance sampling, since we only propose relatively small changes to the sample. The new sample is then accepted in the Markov chain with probability 
\begin{equation}
P_\text{accept} = \min\left( 1, \mathrm{e}^{-S_4(a,b,c,d) + S_4(a',b',c',d')} \right).
\end{equation}

For each $\theta$ we perform 2000 Monte-Carlo runs in parallel with up to $10^6$ samples per run. The results are obtained by averaging over all runs. In the case of small $\theta$, we find in rare cases (one out of 1000 runs) instabilities towards local maxima of the weighting function during the sampling process. This is visible by tracking $S_4(a,b,c,d)$ or the acceptance rate.
This problem can be circumvented by decreasing the size of the update steps, but only at the expense of longer autocorrelation times. As a compromise, we discard runs where $S_4$ falls below a threshold value of $-50$. 

\begin{figure}[t!]
	\centering
	\includegraphics[width=0.45\textwidth]{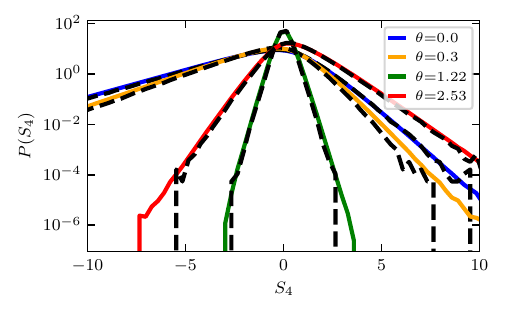}
    \caption{Probability distribution of $S_4(a,b,c,d)$ [Eq.~\eqref{LagrangeG4}] for quasienergy differences $\theta$ as indicated: comparison between ED results (black dashed lines) and MC results (solid coloured lines). Parameters as in Fig.~\ref{fig:synopsis}.
    }
	\label{fig:S4histo}
\end{figure}

To provide an overall test of our form [Eq.~\eqref{JDF}] for the JDF, we show in Fig.~\ref{fig:S4histo} a comparison of the distributions of $S_4(a,b,c,d)$ obtained respectively from ED and from MC sampling of the JDF. The excellent agreement between the two distributions over a range of values for $\theta$ is evidence of the internal consistency of our approach.

%%%%%%%%%%%%%%%%%%%%%%%%%%%%%%%%%%%%%%%%%%%%%%%%%%%%%%%%%%%%%%%%%%%%%%%%%%%%%%%%%%%%%%%%%
\section{Perturbative treatment of Lagrange multipliers}\label{sec:perturbation theory}
%%%%%%%%%%%%%%%%%%%%%%%%%%%%%%%%%%%%%%%%%%%%%%%%%%%%%%%%%%%%%%%%%%%%%%%%%%%%%%%%%%%%%%%%%

An obvious approach to calculations based on the joint eigenstates distributions of Eqns.~\eqref{P2} and \eqref{JDF} is a perturbative expansion in powers of the Lagrange multipliers $G_2(X,\theta)$ and $G_4(X,Y,\theta)$. In this section we set out a general framework for such an expansion and apply it in several ways. While the expansion does not generate fundamentally new results, it provides a useful perspective that is complementary to the one set out in Sec.~\ref{sec:Synopsis} and Sec.~\ref{sec:Implementation}.

We use the expansion to provide an alternative justification of the fitting procedure for $G_4(X,Y,\theta)$ to the one described in Sec.~\ref{G4fit}.  We also use it to consider Eq.~\eqref{JDF} without the simplification employed in Sec.~\ref{subsec2C} of setting $G_2(X,\theta)=0$. We show that for large $q^{L}$ and $q^{L(X)}$ that
if $F_2(X,\theta)$ is calculated from the JDF for four eigenstates rather than two, the influence of $G_4(X,Y,\theta)$ on the result is small. This means that the predictions of ETH for matrix elements between pairs of eigenstates are only weakly affected by the correlations between sets of four eigenstates $F_4(X,Y,\theta)$ that are introduced with our Ansatz for the JDF. Finally, we show that the effect of the Lagrange multipliers $G_2(X,\theta)$ and $G_4(X,Y,\theta)$ on the normalisation of eigenstates drawn from the joint distributions is small in large systems. 

An exact perturbative expansion requires the evaluation of averages over a Haar distribution of vectors, denoted by $P_1^{0)}(a)$, $P_2^{(0)}(a,b)$ and $P_4^{(0)}(a,b,c,d)$ in Sec.~\ref{subsec2B2}. While the formalism required for this is well developed (see e.g. \cite{Brouwer1996Diagrammatic}), it is quite cumbersome. Moreover, we require results only at leading order for $q^L$ large. These can be obtained by substituting in place of the Haar distribution, one in which each vector component is an independent Gaussian random variable. Specifically, consider a computational basis $\{\ket{k_c}\}$
and denote the overlap of the eigenstate $\ket{a}$ with the basis state $\ket{k_c}$ by $a(k)=\braket{k_c|a}$. Define $S_1(a)$ by
\begin{eqnarray}\label{JDF0}
    S_1(a)= q^L \sum_k |a(k)|^2\,.
\end{eqnarray}
We replace $P_4^{(0)}(a_1,a_2,a_3,a_4)$ by % in Eq.~\eqref{JDF} by 
\begin{equation}\label{newnorm}
    P^{(\rm G)}_4(a_1,a_2,a_3,a_4) = \left(\frac{q^L}{\pi}\right)^{4q^L}   e^{-\sum_i S_1(a_i)}\,.
\end{equation}
Although different vectors drawn from this distribution are not in general exactly orthonormal, orthonormality is recovered in the limit $q^L \rightarrow \infty$. We denote averages with respect to this Gaussian distribution by $[\ldots ]_{\rm G}$. 

%%%%%%%%%%%%%%%%%%%%%%%%%%%%%%%%%%%%%%%%%%%%%%%%%%%%%%%%
\subsection{Diagrammatic notation}%%%%%%%%%%%%%%%%%%%
%%%%%%%%%%%%%%%%%%%%%%%%%%%%%%%%%%%%%%%%%%%%%%%%%%%%%%%%

It is useful to employ diagrammatic notation for these Gaussian averages. As in Fig.~\ref{fig:one}, eigenstates are represented by circles and index contractions following from the definition of $M_X(abcd)$ are indicated by solid lines carrying arrows that run from states $|a\rangle$ towards conjugate states  $\langle b|$; these lines carry labels to indicate the subsystem within which the contraction is done. The combination of circles and full lines, with labels for states and subsystems, is fixed by the choice of quantity we average, and diagrams are generated by making all possible Wick pairing of circles. These pairings are represented by dashed lines. The contribution of a diagram is a product of two factors. One factor stems from Eq.~\eqref{newnorm} and consists of $q^{-L}$ for every dashed line. The other factor arises from sums over the Hilbert space at each site. To evaluate this factor we form closed loops in the diagram consisting alternately of dashed lines and full lines traversed in the direction of the arrows. These full lines must carry the label that corresponds to a subsystem in which the site lies. For each site we have a factor of $q$ for every such closed loop.

\begin{figure}[t!]
\begin{center}
\vspace{0.3cm}
\includegraphics[width=8cm]{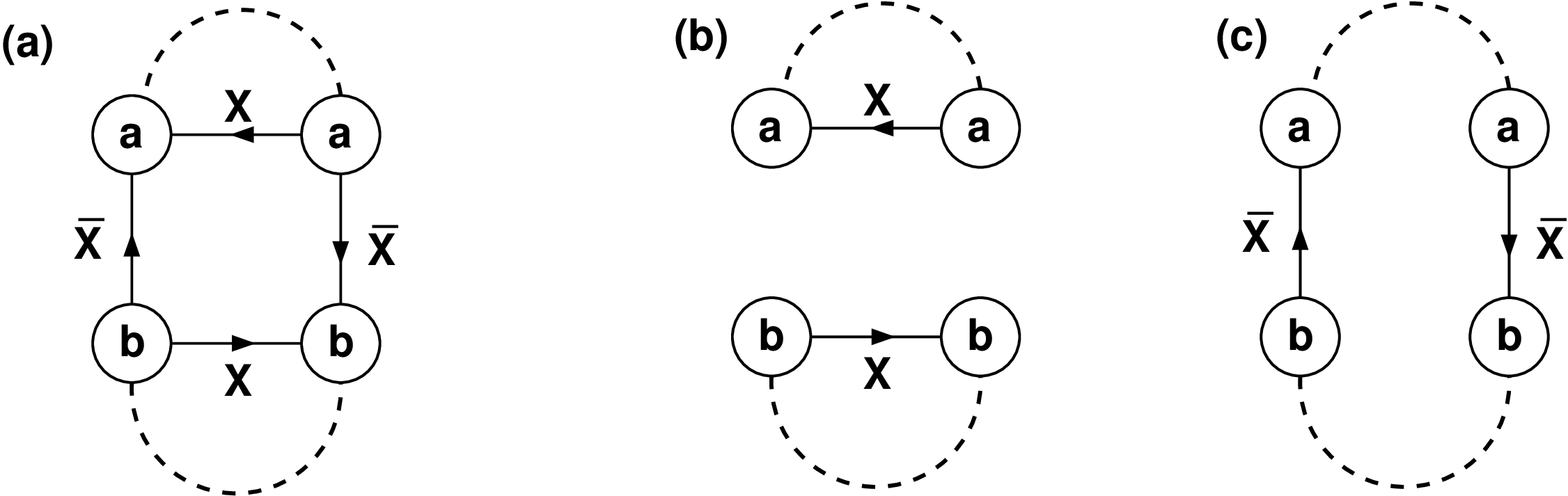}
\end{center}
\caption{Diagrammatic representation and evaluation of $[M_X(abba)]_{\rm G}$: (a) full diagram; (b) decomposition into loops for a site in subsystem $X$; (c) decomposition into loops for a site in subsystem $\overline{X}$.
}\label{figF2}
\end{figure}

For example, the evaluation of $[M_X(abba)]_{\rm G}$ is shown in Fig.~\ref{figF2}.
It yields
\be\label{MG}
[M_X(abba)]_{\rm G} = q^{-2L} q^{2L(X)} q^{L(\overline{X})} = q^{L(X) - L}\,.
\ee

%%%%%%%%%%%%%%%%%%%%%%%%%%%%%%%%%%%%%%%%%%%%%%%%%%%%%%%%%%%%%%%%%%%%%%%%%%%%%%%%%%%%%%%%%%%
\subsection{Relating $F_4(X,Y,\theta)$ and $G_4(X,Y,\theta)$}\label{sec:relating}%%%%%%%%%%
%%%%%%%%%%%%%%%%%%%%%%%%%%%%%%%%%%%%%%%%%%%%%%%%%%%%%%%%%%%%%%%%%%%%%%%%%%%%%%%%%%%%%%%%%%%

\begin{widetext}

\begin{figure}[thbp]
\begin{center}
\includegraphics[width=14cm]{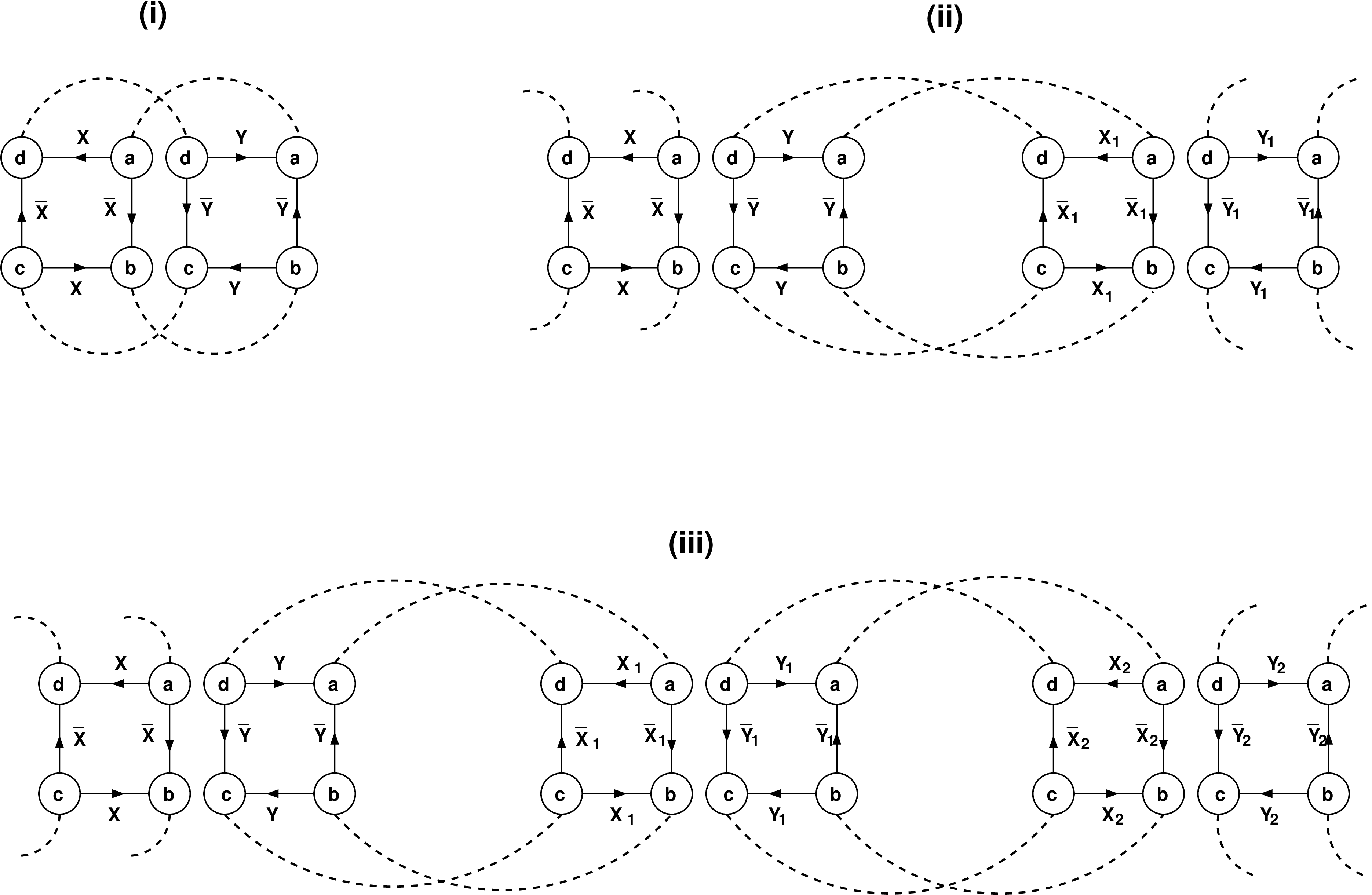}
\end{center}
\caption{Leading connected contributions to $\big[[\mathsf{M}]_X[\mathsf{M}]_Y [ \mathsf{M}^{\rm T}\mathsf{G}_4\mathsf{M}^*]^n\big]_{\rm G}$ for: (i) $n=0$, (ii) $n=1$ and (iii) $n=2$. The pair of squares in (i) and the left-most pairs of squares in (ii) and (iii) represent $[\mathsf{M}]_X[\mathsf{M}]_Y$; other pairs of squares represent $\mathsf{M}^{\rm T}\mathsf{G}_4\mathsf{M}^*$. The dashed lines in (ii) and (iii) that leave the diagrams on the left are joined to the dashed lines that leave on the right. 
}\label{fig:expansion}
\end{figure}

\end{widetext}

Our first objective is to obtain a relationship between $F_4(X,Y,\theta)$ and $G_4(X,Y,\theta)$ by expanding $e^{-S_4(a,b,c,d)}$ in a power series, averaging with respect to the Gaussian distribution, and then resumming the terms at each order in the expansion that are leading for $q^L$ and $q^{L(X)}$ large. By this means we will recover Eq.~\eqref{eq:g}. 

Using the notation of Eq.~\eqref{eq:matrix version} we require the connected contributions to $\big[[\mathsf{M}]_X[\mathsf{M}]_Y [ \mathsf{M}^{\rm T}\mathsf{G}_4\mathsf{M}^*]^n\big]_{\rm G}$  at each order $n$ in perturbation theory that are leading for large system and subsystem sizes. These come from the Wick contractions that generate the largest number of loops in a decomposition of the type illustrated in Fig.~\ref{figF2}. These contractions are illustrated for $n=0, 1$ and $2$ in Fig.~\ref{fig:expansion}, establishing an obvious pattern for general $n$. Retaining only these terms we have
\begin{eqnarray}\label{eq:galt}
    \big[[\mathsf{M}]_X [\mathsf{M}^*]_Y \big]_{\rm av} &\approx& \sum_{n=0}^\infty(-1)^n [(\mathsf{G}^{(0)}_4)^{-1}(\mathsf{G}_4(\mathsf{G}^{(0)}_4)^{-1})^n]_{X,Y}\nonumber\\
    &=& \big[(\mathsf{G}^{(0)}_4)^{-1}[\openone + \mathsf{G}_4(\mathsf{G}^{(0)}_4)^{-1}]^{-1}\big]_{X,Y} \nonumber \\
    &=& \big[ (\mathsf{G}_4 + \mathsf{G}^{(0)}_4  )^{-1} \big]_{X,Y}
\end{eqnarray}
where $\mathsf{G}^{(0)}_4$ is as defined following Eq.~\eqref{eq:new distribution function} by the relation
$[(\mathsf{G_4^{(0)}})^{-1}]_{X,Y}=\big[[\mathsf{M}]_X [\mathsf{M}^*]_Y \big]_{\rm G} = q^{L(X,Y) -4L}$.
For averaging over an Gaussian ensemble, the last relation is exact.
[Note that for these terms the factor of $(n!)^{-1}$ arising from the power series expansion of $e^{-S_4(a,b,c,d)}$ is cancelled by a combinatorial factor arising in the pairing of terms in Fig.~\ref{fig:expansion}.] Comparing Eqns.~\eqref{eq:g} and \eqref{eq:galt} we see that this diagrammatic resummation provides an alternative derivation of the main results of Sec.~\ref{G4fit}.

%%%%%%%%%%%%%%%%%%%%%%%%%%%%%%%%%%%%%%%%%%%%%%%%%%%%%%%%%%%%%%%%%%%%%%%%
\subsection{Cross-correlations}\label{sec:crosscorrelations}%%%%%%%%%
%%%%%%%%%%%%%%%%%%%%%%%%%%%%%%%%%%%%%%%%%%%%%%%%%%%%%%%%%%%%%%%%%%%%%%%%

In our MC studies of the eigenstate JDF [Eq.~\eqref{JDF}] we have made two simplifications: one is to omit $G_4(X,Y,\theta)$ when studying $F_2(X,\theta)$, and the other is to omit $G_2(X,\theta)$ when studying $F_4(X,Y,\theta)$. The magnitudes of the resulting errors can be assessed using perturbation theory, as we now discuss. 

\begin{figure}[thbp]
\begin{center}
\includegraphics[width=7cm]{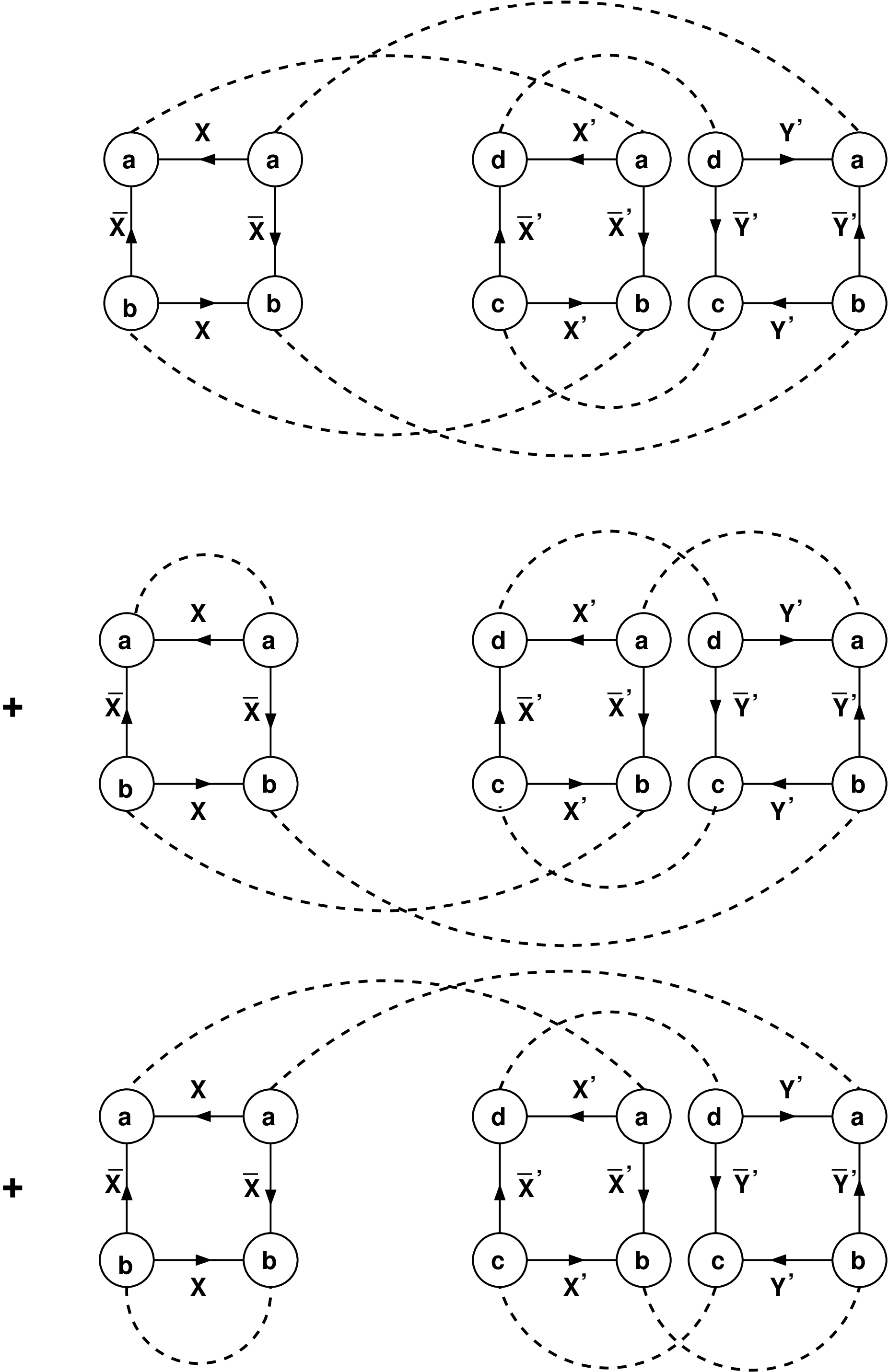}
\end{center}
\caption{Contributions to the connected average
%$[M_X(abba)M_{X^\prime}(abcd)M_{Y^\prime}^*(abcd)]_{\rm G,c}$, as 
defined in Eq.~\eqref{eq:connected}
}\label{fig:connected}
\end{figure}

We begin by considering the effect of $G_4(X,Y,\theta)$ on the value of $F_2(X,\theta)$, or equivalently on the value of $[M_X(abba)]_{\rm av}$, as follows. To first order in perturbation theory in $G_4(X,Y,\theta)$ we have
\begin{eqnarray}\label{eq:expansion}
    [M_X(abba)]_{\rm av} = && [M_X(abba)]_{\rm G}  \nonumber \\
    -\sum_{X^\prime Y^\prime} G_4(X^\prime,Y^\prime,&&\theta)[M_{X}(abba)M_{X^\prime}(abcd)M^*_{Y^\prime}(abcd)]_{G,c}\,, \nonumber \\
    && + {\cal O}([G_4(X^\prime,Y^\prime,\theta]^2)\,,
\end{eqnarray}
where $[\ldots]_{\rm G,c}$ denotes the connected average, defined by
\begin{eqnarray}\label{eq:connected}
   [M_X(abba)M_{X^\prime}(abcd)  M_{Y^\prime}^*(abcd)  ]_{\rm G,c}  = && \nonumber \\
{ [ M_X(abba)  M_{X^\prime}(abcd)  M_{Y^\prime}^*(abcd)  ]_{\rm G} } &&  \nonumber\\
   -  {[M_X(abba)]_{\rm G}[M_{X^\prime}(abcd) M_{Y^\prime}^*(abcd) ]_{\rm G}} && \,.
\end{eqnarray}
The diagrams that contribute to this connected average are shown in Fig.~\ref{fig:connected}. Evaluating these diagrams for the representative case $X = X^\prime = Y^\prime$, we obtain
\begin{eqnarray}\label{eq:conn}
[M_X(abba)&&M_{X}(abcd)  M_{X}^*(abcd)  ]_{\rm G,c} \nonumber \\
= q^{-3L}&&(q^{-L(X)}+ 2q^{-L(\overline{X})})\,.
\end{eqnarray}
We compare the zeroth-order term, which is 
\begin{equation}
[M_X(abba)]_{\rm G} = q^{L(X) - L}
\end{equation}
from Eq.~\eqref{MG}, with the first-order term, using Eq.\eqref{eq:conn} and $G_4(X,X,\theta) \sim q^{2L}$ from Eq.~\eqref{eq:new distribution function}, which gives
\begin{eqnarray}\label{pertG4}
    \sum_{X^\prime Y^\prime} G_4(X^\prime,Y^\prime&&,\theta)[M_{X}(abba)M_{X^\prime}(abcd)M^*_{Y^\prime}(abcd)]_{G,c}\,, \nonumber \\
    \sim&&  q^{-L}(q^{-L(X)} + 2 q^{-L(\overline{X})})\,.
\end{eqnarray}
Hence we see that the effect of $G_4(X^\prime,Y^\prime,\theta)$ on $F_2(X,\theta)$ is small provided that $q^{L(X)}$ and $q^{L(\overline{X})}$ are large.

We have not systematically investigated higher order terms in Eq.~\eqref{eq:expansion}, but we expect them individually to be small: note that although $G_4(X^\prime,Y^\prime,\theta)\sim q^{2L}$, each such contribution is accompanied by a factor of $q^{-4L}$ from extra dashed lines, as well as diagram-dependent factors from sums over the Hilbert space at each site.

A similar study of the effect of $G_2(X,\theta)$ at first order in perturbation theory on $F_4(X,Y,\theta)$, or equivalently on $[M_{X^\prime}(abcd)M_{Y^\prime}^*(abcd)]_{\rm av}$ reaches a different conclusion: the influence is not small in $q^L$ or $q^{L(X)}$, but only in powers of $G_2(X,\theta)$. This is not unexpected: the functional form for a correlator involving four eigenstates and depending on three quasienergy differences has been discussed previously \cite{Chan2019Eigenstate} and in Sec.~\ref{subsec:multi}; it involves factors of both $F_2(X,\theta)$ and $F_4(X,Y,\theta)$ (see Eq.~\eqref{eq:multi}). We do not pursue this further since we have chosen here to study models in which $F_2(X,\theta)$ lies close to its value for Haar-distributed pairs of eigenstates, and $G_2(X,\theta)$ is therefore small.

%%%%%%%%%%%%%%%%%%%%%%%%%%%%%%%%%%%%%%%%%%%%%%%%%%%%%%%%%%%%%%%%%%%%%
\subsection{Renormalisation of propagators and vertices}%%%%%%%%%%
%%%%%%%%%%%%%%%%%%%%%%%%%%%%%%%%%%%%%%%%%%%%%%%%%%%%%%%%%%%%%%%%%%%%%

\begin{figure}[thbp]
\begin{center}
\includegraphics[width=6.5cm]{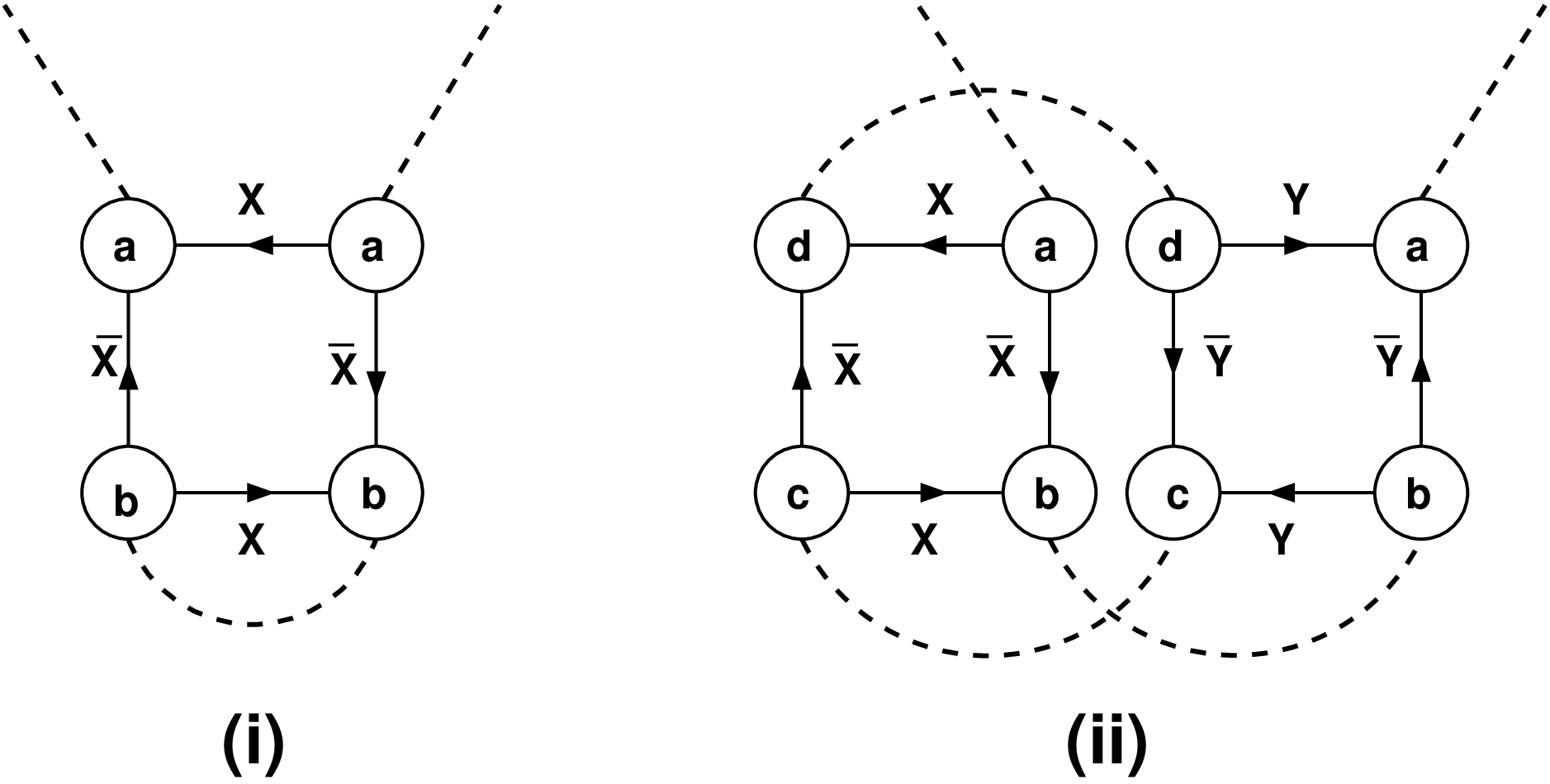}
\end{center}
\caption{Contributions to the self-energy $\Sigma$ at first order: (i) from $G_2(X,\theta)$, and (ii) from $G_4(X,Y,\theta)$. 
}\label{fig:selfenergy}
\end{figure}

An alternative perspective is provided by considering the perturbative renormalisation of the propagators and vertices appearing in the JDF. We begin with the former: the bare propagator in the theory, represented using dashed lines in the figures, is generated by Eqns.~\eqref{JDF0} and \eqref{newnorm}, and carries a factor of $q^{-L}$. More generally, it acquires a self-energy $\Sigma_{}$
%, which is a diagonal matrix in the computational basis introduced in Eq.~\eqref{JDF0}, 
and the factor becomes $(q^L + \Sigma_{})^{-1}$. Our aim is to evaluate $\Sigma_{}$ at leading order in $G_2(X,\theta)$ and $G_4(X,Y,\theta)$ and compare it to the bare inverse propagator $q^L$: if it is small under this comparison, then the effects of the vertices on the normalisation of eigenstates can be neglected. The diagrams contributing to $\Sigma$ at this order are shown in Fig.~\ref{fig:selfenergy}. They are diagonal matrices in the site basis and the magnitudes of their largest entries are respectively 
\begin{equation}\label{Sigma(i)}
    \Sigma^{\rm (i)}_{kk} = q^{-L} \sum_X G_2(X,\theta) q^{L(X)}  \sim q^{2L(X)}
\end{equation}
[where we have used Eqns.~\eqref{G} and \eqref{M0} to estimate $G_2(X,\theta) \sim q^{L+L(X)}$] and
\begin{equation}\label{Sigma(ii)}
    \Sigma^{\rm (ii)}_{kk} = q^{-4L}\sum_{X,Y} q^{L(X,Y)}G_4(X,Y,\theta) \sim {\cal O}(1)\,,
\end{equation}
[where we have used an estimate of $G_4(X,Y,\theta)$ given above Eq.~\eqref{pertG4}]. 
From this we see that the renormalisation of the propagator is indeed small if $q^L$ is large and if $q^{2L(X)}\ll q^L$. In the opposite regime ($q^{2L(X)} > q^L$) we believe that $G_2(X,\theta)$ is small since $F_2(X,\theta)$ approaches the value it takes for Haar-distributed eigenstates, and that $\Sigma^{\rm (i)}_{kk} \ll q^L$ notwithstanding the estimate of Eq.~\ref{Sigma(i)}.

\begin{figure}[thbp]
\begin{center}
\includegraphics[width=6cm]{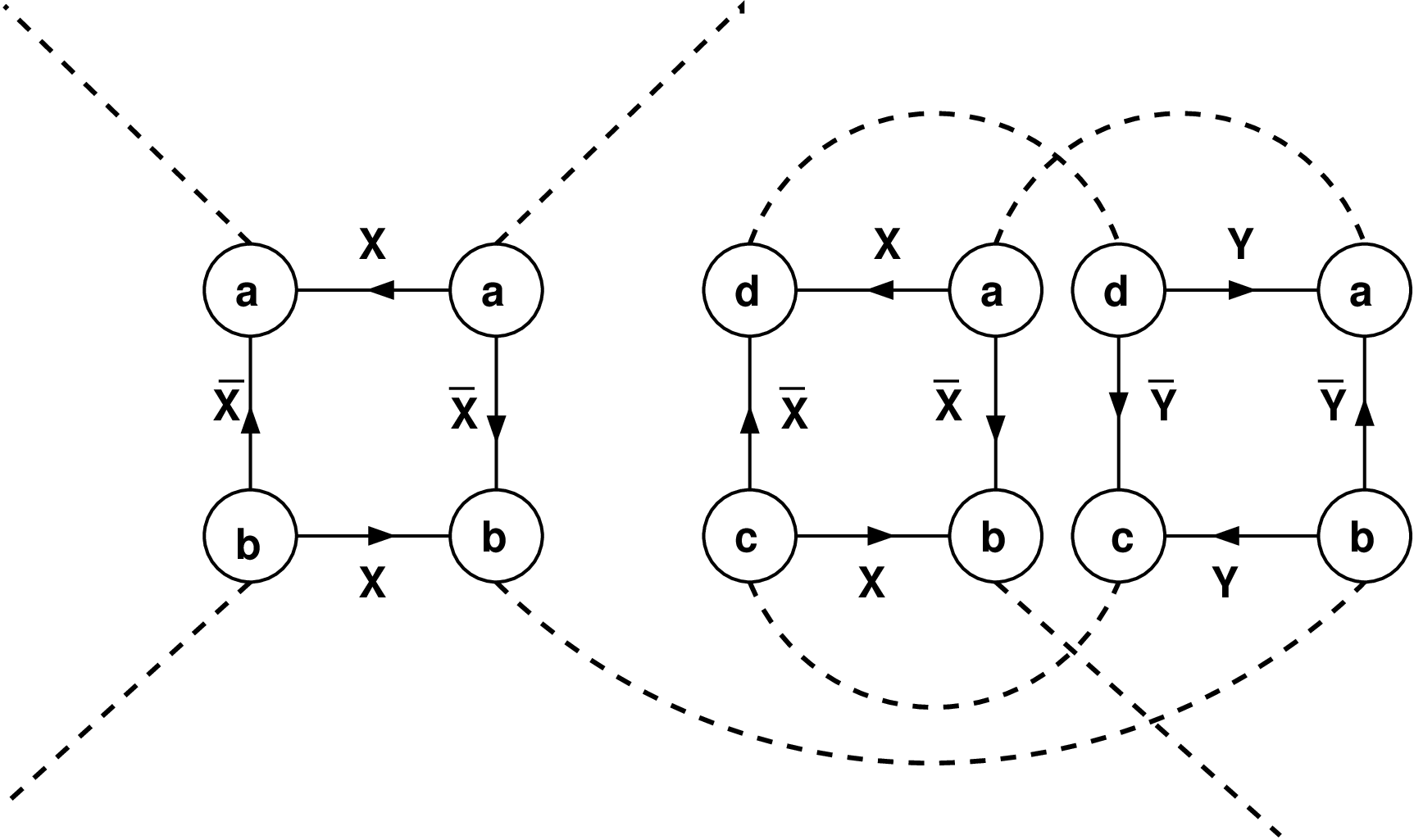}
\end{center}
\caption{One of three contributions to the renormalisation at first order in both $G_2(X,\theta)$ and in $G_4(X,Y,\theta)$ of the vertex that appears in $S_2(a,b)$. 
}\label{fig:renormalisedvertex}
\end{figure}

A similar discussion can be developed of renormalised vertices generated by combining contributions from $S_2(a,b)$ and $S_4(a,b,c,d)$ in \eqref{JDF} and forming sufficient internal contractions to generate a new effective contribution to either $S_2(a,b)$ or $S_4(a,b,c,d)$. For example, one (out of three possible terms) contributing in this way to $S_2(a,b)$ at first order in both $G_2(X,\theta)$ and $G_4(X,Y,\theta)$ is shown in Fig.~\ref{fig:renormalisedvertex}. We do not pursue this further because this is exactly the same phenomenon as has been discussed from a different perspective in Sec.~\ref{sec:crosscorrelations}. Note that it is not necessary to consider renormalisation of $S_2(a,b)$ in powers of $G_2(X,\theta)$ alone, or of $S_4(a,b,c,d)$ in powers of $G_4(X,Y,\theta)$ alone, since these effects (which are not generally small) are covered in full by the approaches described in Sec.~\ref{sec:relating} and in Sec.~\ref{sec:probfunction}.

%%%%%%%%%%%%%%%%%%%%%%%%%%%%%%%%%%%%%%%%%%%%%%%%%%%%%%%%%%%%%%%%%%%%%%%%
\section{Results for further models}\label{sec:further models}%%%%%%%%%%
%%%%%%%%%%%%%%%%%%%%%%%%%%%%%%%%%%%%%%%%%%%%%%%%%%%%%%%%%%%%%%%%%%%%%%%%
In this section we provide numerical results for brickwork models additional to the one treated in Sec.~\ref{sec:Synopsis}. 
In Sec.~\ref{sec:q2Haarresults} we omit the cutoff in the operator purity of gates introduced in Sec.~\ref{subsec:Models} and present results for gates drawn from a Haar-distribution.
In Sec.~\ref{sec:q3results} we give results with local Hilbert space dimension $q=3$ and Haar-distributed gates.

%%%%%%%%%%%%%%%%%%%%%%%%%%%%%%%%%%%%%%%%%%%%%%%%%%%%%%%%%%%%%%%%%%%%%%%%%%%%%%%%%%
\subsection{Results for $q=2$ with Haar gates}~\label{sec:q2Haarresults}%%%%%%%%%%
%%%%%%%%%%%%%%%%%%%%%%%%%%%%%%%%%%%%%%%%%%%%%%%%%%%%%%%%%%%%%%%%%%%%%%%%
 \begin{figure}[t]
	\centering
	\includegraphics[width=0.48\textwidth]{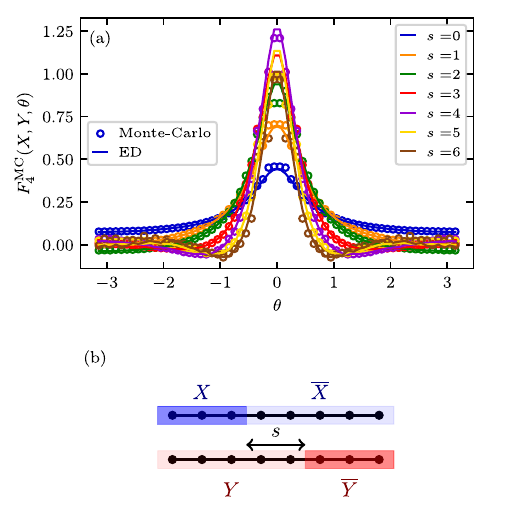}
	\caption{$F^{\rm MC}_4(X,Y,\theta)$ (open circles from MC) and $F_4(X,Y,\theta)$ (lines from ED) vs $\theta$ for various $s$, for a brickwork model with Haar-distributed gates, $q=2$ and $L=8$. 
 %The agreement is up to small angles excellent.
	\label{fig:CorrelationsL8q2Haar}}
\end{figure}
As discussed in Sec.~\ref{sec:Implementation}, the determination of the Lagrange multipliers $G_4(X,Y,\theta)$ in the brickwork model with Haar-distributed gates is complicated by the presence of weak links. Specifically, we find that the simplifying assumption used in Sec.~\ref{G4fit}, namely that the probability distribution of $M_X(abcd)$ is approximately Gaussian, does not hold for Haar-distributed gates. Instead, this distribution exhibits long tails, as we demonstrate in Appendix~\ref{subsec:Weak links}.

To determine $G_4(X,Y,\theta)$ under these circumstances, we use the iterative fitting procedure introduced in Sec.~\ref{iter}, which is not predicated on a particular form for the probability distribution of $M_X(abcd)$. We take all subsystems $X$ and $Y$ that can be obtained using a single cut from a system with open boundary conditions. In contrast to Sec.~\ref{sec:Synopsis}, where the requirement of an approximately Gaussian distribution for $M_X(abcd)$ led to the restriction $2 < L(X) < L-2$, here we include all subsystem sizes $1\leq L(X) \leq L-1$. A disadvantage of the iterative fitting procedure is that it requires multiple MC evaluations of $F_4^{\rm MC}(X,Y,\theta)$, which is slow if the total number of degrees of freedom $q^L$ involved is large; this restricts us to $L=8$ with $q=2$. In this case estimates $F^{\rm MC}_4(X,Y,\theta)$ are obtained using between 60 and 300 iterations and a step size $\alpha=0.2$. The MC results shown in Fig.~\ref{fig:CorrelationsL8q2Haar} display very good agreement with ED data.
%across all values of the quasienergy difference $\theta$. 

%%%%%%%%%%%%%%%%%%%%%%%%%%%%%%%%%%%%%%%%%%%%%%%%%%%%%%%%%%%%%%%%%%%%%%%%%%%%%%%%%%%%%%%%%%%%%%%%%%
\subsection{Results for \texorpdfstring{$q=3$}{q=3} brickwork model}~\label{sec:q3results}%%%%%%%%

%%%%%%%%%%%%%%%%%%%%%%%%%%%%%%%%%%%%%%%%%%%%%%%%%%%%%%%%%%%%%%%%%%%%%%%%%%%%%%%%%%
 \begin{figure}[t]
	\centering
    \includegraphics[width=0.48\textwidth]{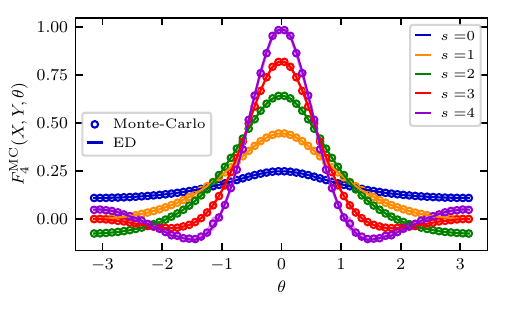}
	\caption{$F^{\rm MC}_4(X,Y,\theta)$ (open circles from MC) and $F_4(X,Y,\theta)$ (lines from ED) vs $\theta$ for various $s$, for a brickwork model with Haar-distributed gates, $q=3$ and $L=8$. 
	\label{fig:Correlationsq3}}
\end{figure}

\begin{figure}[t]
\includegraphics[width=0.48\textwidth]{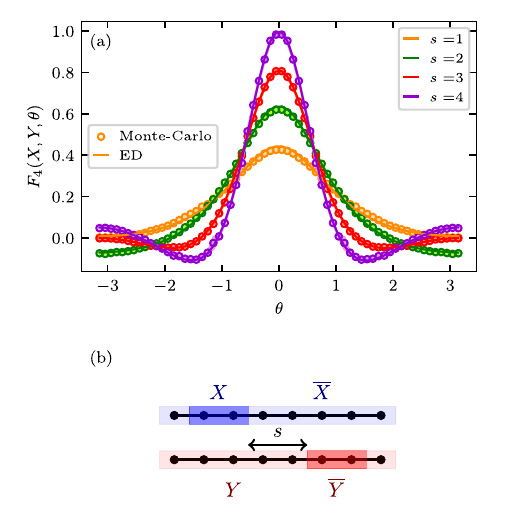}
\caption{Test of JDF fitted to behaviour in the geometries of Fig.~\ref{fig:Correlationsq3}~(b) but applied to geometries of Fig.~\ref{fig:q3resultsOtocs}~(b), for the brickwork model with $q=3$ and $L=8$. (a) Comparison of data from MC (open circles) and ED (solid lines). 
(b) Partition used for (a), in which the 8-site system is divided by two spatial cuts into a two-site subsystem $X$ and its complement $\overline{X}$, or a two-site subsystem $\overline{Y}$ and its complement $Y$. 
}\label{fig:q3resultsOtocs}
\end{figure}

In our study of the brickwork model with $q=3$ and Haar-distributed gates we use system size $L=8$ so that ED calculations are straightforward.
The effect of weak links becomes less pronounced with increasing bond dimension $q$ and as a consequence the single-shot approach to obtain $G_4(X,Y,\theta)$ of Sec.~\ref{G4fit} becomes more accurate. Conversely, for a given system size the iterative method of Sec.~\ref{iter} is more difficult to apply with increasing $q$, because a large number of samples is required to obtain accurate estimates for $F^{\rm MC}_4(X,Y,\theta)$ when the Hilbert space dimension $q^L$ of the system is large (see the discussion of Sec.~\ref{subsec2B} and Ref.~\cite{chan_eigenstate_2019}).  
We therefore determine Lagrange multipliers for this model using the single-shot approach, taking subsystems $X$ obtained from a system with open boundary conditions by making a single cut, and with $L(X)\geq 2$, $L(\overline{X})\geq 2$. 
The MC results shown in Fig.~\ref{fig:Correlationsq3} display excellent agreement with ED data.

Furthermore, we test how well the JDF fitted to the geometries of Fig.~\ref{fig:CorrelationsL8q2Haar}(b) can reproduce the behaviour of the OTOC in the geometries of Fig.~\ref{fig:q3resultsOtocs}(b). The results shown in Fig.~\ref{fig:q3resultsOtocs}(a) display excellent agreement between ED and MC data.

\subsection{Results for \texorpdfstring{$q=2$}{q=2} brickwork model with periodic boundary conditions}~\label{sec:q2resultspbc}%%%%%%%%

%%%%%%%%%%%%%%%%%%%%%%%%%%%%%%%%%%%%%%%%%%%%%%%%%%%%%%%%%%%%%%%%%%%%%%%%%%%%%%%%%%
 \begin{figure}[t]
	\centering
    \includegraphics[width=0.48\textwidth]{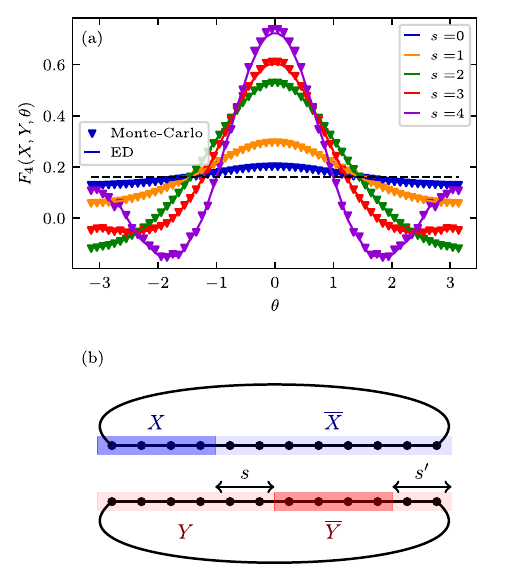}
	\caption{$F^{\rm MC}_4(X,Y,\theta)$ (open circles and solid triangles from MC) and $F_4(X,Y,\theta)$ (lines from ED) vs $\theta$ for various $s$, for a brickwork model with Haar-distributed gates and periodic boundary conditions, $q=2$ and $L=12$ for the geometries in displayed in~(b). Partitions in the Lagrange multipliers are generated using two cuts (solid triangles). The results generated from Monte-Carlo sampling show very good agreement with the data obtained from exact diagonalization. (b) Partition used for (a). Due to the periodic boundary conditions, two separations $s$ and $s'$ between the partitions exist. We choose $s=s'$.
	\label{fig:Correlationspbc}}
\end{figure}

Finally, we show results for the brickwork model for $q=2$, $L=12$ using periodic boundary conditions.
This leads to a second possible direction for the growth of correlations and two separations between the partitions denoted by $s$ and $s'$, with the most important separation being the minimum of the two. To simplify the presentation we set $s=s'$.

The results are shown in Fig.~\ref{fig:Correlationspbc}. 
In this case, we have used the exact Haar average $\mathsf{G}_4^{(0)}$ instead of the leading approximation given in Eq.~\eqref{HaarCovariance}. 
The solid triangles indicate $F^{\rm MC}_4(X,Y,\theta)$ calculated Monte-Carlo sampling from a JDF that includes Lagrange multipliers $G_4(X,Y,\theta)$ for all connected partitions $X$, $Y$ generated by two cuts with $L(X), L(Y) >1$.
The results in this case~(open circles) show very good agreement with $F_4(X,Y,\theta)$ calculated using ED.

%%%%%%%%%%%%%%%%%%%%%%%%%%%%%%%%%%%%%%%%%%%%%%%%%%%%%%%%%%%%%%%%%%%%%%%%%%%%
\section{Summary and outlook \label{sec:Discussion and outlook}}%%%%%%%%%
%%%%%%%%%%%%%%%%%%%%%%%%%%%%%%%%%%%%%%%%%%%%%%%%%%%%%%%%%%%%%%%%%%%%%%%%%%%%

In this work we have analysed the interplay between the statistical properties of the time evolution operator for chaotic many-body quantum systems, and the quantum information dynamics that these systems display. The eigenstate thermalisation hypothesis provides an accurate description of some key aspects, in terms of the probability distribution of matrix elements of observables between eigenstates of the time-evolution operator. In its original form, however, it does not capture the consequences of a finite speed for quantum information spreading. To remedy this, we advocate a change of viewpoint. In place of the matrix elements considered within ETH, we construct correlators from eigenstates. We show that the simplest such correlator capturing signatures of quantum information dynamics at long times and large distances is unique and involves a set of four eigenstates. We also propose an Ansatz for the joint probability distribution function of small numbers of eigenstates, in which the values of correlators are controlled by Lagrange multipliers. We support these general ideas using numerical studies of Floquet quantum circuits, showing firstly that correlators have the expected features, and secondly that it is possible to choose values for the Lagrange multipliers so that our Ansatz reproduces these features accurately. We believe our approach is complementary to recent work \cite{pappalardi_eigenstate_2022,pappalardi_general_2023} that generalises ETH using the language of Free Probability theory. 

An advantage of viewing this problem in terms of correlations between eigenstates rather than matrix elements is that even the simplest assumption, of a Haar distribution for eigenstates, yields correlations with the correct order of magnitude. In this instance, however, the correlations are independent of eigenvalues. By contrast, a microscopic model for local quantum dynamics yields a characteristic dependence of correlators on differences in quasienergies, as we have summarised in Fig.~\ref{fig:synopsis}. We have shown that our Ansatz for the joint distribution of eigenstates, together with a suitable choice for the Lagrange multipliers, captures this dependence on eigenvalues differences. 

The quantity that characterises correlations between sets of four eigenstates, denoted $F_4(X,Y,\theta)$ in this paper, is a function not only of the eigenphase difference $\theta$, but also of subsystem choices $X$ and $Y$. This constitutes a large set of possibilities if there are no restrictions on the way the subsystems are selected. We have focussed on the simple set of subsystems that can be obtained from an overall system with open (rather than periodic) boundary conditions by making a single spatial cut, choosing Lagrange multipliers in our Ansatz for the JDF so that $F_4(X,Y,\theta)$ is reproduced accurately for these subsystems. In addition we have shown that the eigenstate correlations imposed in this way are sufficient to reproduce to a good approximation the correlators for some other choices of $X$ and $Y$. In particular, we have demonstrated that the behaviour of the OTOC for operators supported on a few sites of the system (requiring choices of $X$ and $Y$ each involving two cuts) is well approximated by our JDF. 

Several obvious directions remain open for future work. Perhaps most importantly, while our discussion in this paper has been restricted to Floquet systems, it would be desirable to extend the approach to systems with a time-independent Hamiltonian. In our context, the significance of such an extension is that diagonal matrix elements of local observables in the basis of Hamiltonian eigenstates are generically functions of energy, a feature absent from Floquet systems. Indeed, ETH is formulated in part to describe this energy dependence. A natural  modification of our JDF to impose such an energy dependence is to supplement Eqns.~\eqref{P2} and \eqref{JDF} with an extra factor $e^{-\sum_kS_1(a_k)}$ chosen to bias selected eigenstates towards a pre-specified energy shell. An obvious choice is to take $S_1(a)= \beta \langle a|H|a\rangle$, where $H$ is the Hamiltonian and $\beta$ is a Lagrange multiplier. 

It is worth emphasising that the correlations induced respectively by $S_1(a)$, $S_2(a,b)$ and $S_4(a,b,c,d)$ are associated with widely separated energy scales. Taking the characteristic strength of local interactions as the unit of energy, $S_1(a)$ selects vectors from a broad energy window, which has a width that increases with system size.
%of order $L^{1/2}/\beta$. 
In turn, $S_2(a,b)$ generates correlations between pairs of vectors within an energy window with width of order unity. Finally $S_4(a,b,c,d)$ generates correlations between groups of four vectors with energy or quasienergy differences lying in a narrow energy window, whose width decreases indefinitely with increasing separation between the subsystems $X$ and $Y$ in $F_4(X,Y,\theta)$.

A further direction for future work is to examine correlations between sets of $n$ eigenstates with $n>4$. This opens up many possibilities, since generalisations of the correlators $F_2(X,\theta)$ and $F_4(X,Y,\theta)$ may involve multiple subsystems in their definition. Restricting to a pair of subsystems $X$ and $Y$, these higher-order correlators are required, for example, to describe the higher-order R\'enyi entropies of the operator entanglement for the time-evolution operator. Beyond this, one can ask whether there are physical phenomena exposed only by higher-order correlators.

%%%%%%%%%%%%%%%%%%%%%%%%%%%
\section*{Acknowledgments}
%%%%%%%%%%%%%%%%%%%%%%%%%%%

This work was supported by the Deutsche Forschungsgemeinschaft through the cluster of excellence ML4Q (EXC2004, project-id 390534769) and by the UK Engineering and Physical Sciences Research Council through Grants EP/N01930X/1  and EP/S020527/1. We also acknowledge support from the QuantERA II Programme, which has received funding from the European Union’s Horizon 2020 research innovation programme (GA 101017733), and from the DFG through the project DQUANT (project-id 499347025), and from the National Science Foundation under Grant No. NSF PHY-1748958. We thank S. Parameswaran for useful comments.
\appendix
\section{Distribution of $M_X(abcd)$}\label{sec:Finitesizeeffects}
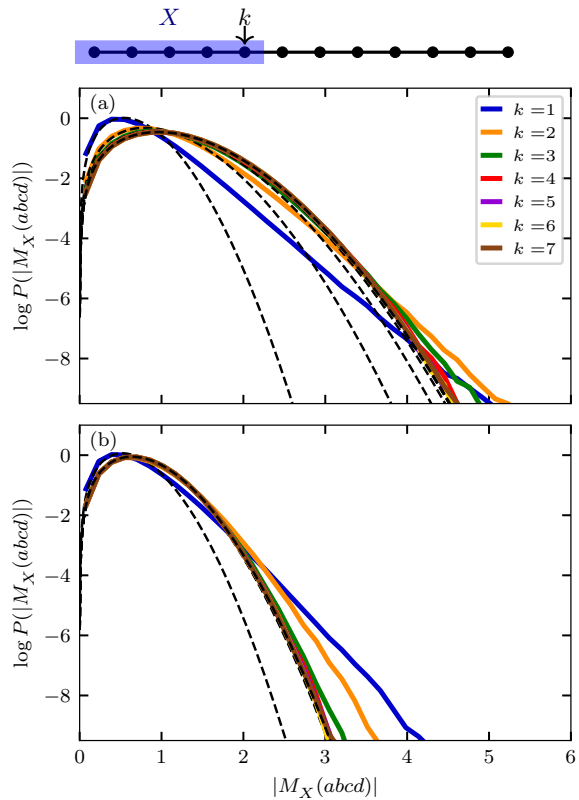
\begin{figure}[h!]
	\centering
    %\documentclass{standalone}
%\usepackage{tikz}
%\begin{document}
        \begin{tikzpicture}
        \draw[very thick] (0.5,0) -- (6,0);
        \foreach \x in {0.5,1.0,1.5,2.0,2.5,3.0,3.5,4.0,4.5,5.0,5.5,6.0}
            \filldraw (\x,0) circle (2pt);
            \filldraw[opacity=0.4,color=blue] (0.25,-0.15) rectangle (2.75,0.15);
            %\filldraw[opacity=0.4,color=red] (3.75,-0.15) rectangle (6.25,0.15);
            %\draw[very thick,<->] (2.75,0.25) -- node[above] {$s$} (3.75,0.25);
            \node[anchor=south,color=blue!50!black] (x) at (1.5,0.25) {$X$};
\node[anchor=south,color=blue!0!black] (k) at (2.5,0.25) {$k$};
\draw[thick][->] (2.5,0.35) to     (2.5,0.1);
            
            %\node[anchor=south,color=red!50!black] (y) at (5,0.25) {$Y$};
    \end{tikzpicture}
%\end{document}        
    \vspace{0.4cm}
	\includegraphics[width=0.45\textwidth]{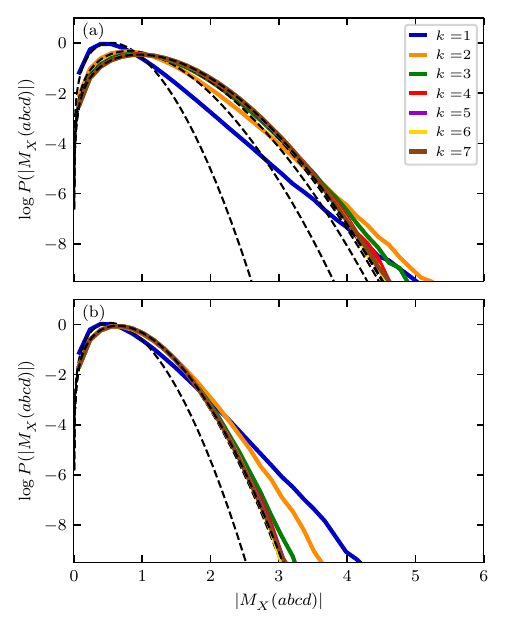}
    \caption{Probability distribution of $|M_X(abcd)|$ for different subsystems $X$ with $L(X)=k$ and $L=12$, $q=2$, and fitted complex Gaussian distributions~(dashed black) as guide for the eye: (a) $\theta=0.1$ and (b) $\theta=1.7$. Brickwork model with gates sampled from the distribution defined in ~\ref{subsec:Models} and a cutoff $0.3 \times q^4$ for the operator purity. 
    With decreasing partition size $L(X)$, the distribution of $M_X(abcd)$ exhibits non-Gaussian tails. }
	\label{fig:M0histodistribution}
\end{figure}
\begin{figure}[h!]
	\centering
    %\documentclass{standalone}
%\usepackage{tikz}
%\begin{document}
        \begin{tikzpicture}
        \draw[very thick] (0.5,0) -- (6,0);
        \foreach \x in {0.5,1.0,1.5,2.0,2.5,3.0,3.5,4.0,4.5,5.0,5.5,6.0}
            \filldraw (\x,0) circle (2pt);
            \filldraw[opacity=0.4,color=blue] (0.25,-0.15) rectangle (2.75,0.15);
            %\filldraw[opacity=0.4,color=red] (3.75,-0.15) rectangle (6.25,0.15);
            %\draw[very thick,<->] (2.75,0.25) -- node[above] {$s$} (3.75,0.25);
            \node[anchor=south,color=blue!50!black] (x) at (1.5,0.25) {$X$};
\node[anchor=south,color=blue!0!black] (k) at (2.5,0.25) {$k$};
\draw[thick][->] (2.5,0.35) to     (2.5,0.1);
            
            %\node[anchor=south,color=red!50!black] (y) at (5,0.25) {$Y$};
    \end{tikzpicture}
%\end{document}        
    \vspace{0.4cm}
	\includegraphics[width=0.45\textwidth]{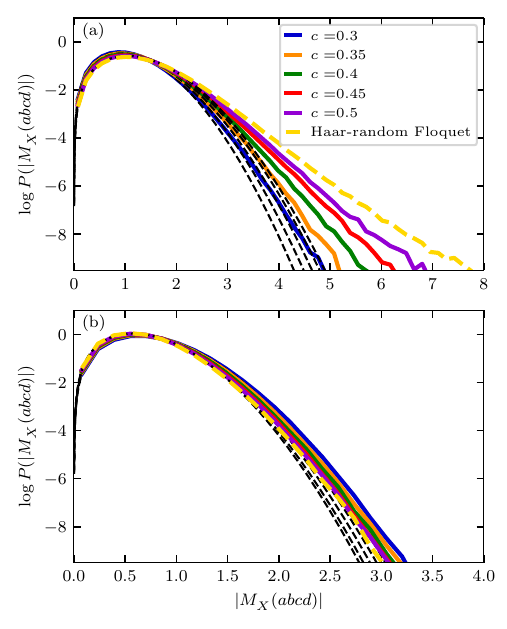}
    \caption{Probability distribution of $|M_X(abcd)|$ at partition $k=2$ and $L=12$, $q=2$, and a complex Gaussian distribution~(dashed black) as guide for the eye: (a) $\theta=0.1$ and (b) $\theta=1.7$. Brickwork model with unitary gates sampled from the distribution defined in ~\ref{subsec:Models} and values of the cutoff $c$ as indicated. 
    With increasing cutoff, the distributions show  non-Gaussian tails for small relative phases $\theta$. 
    }
	\label{fig:M0histodistribution2}
\end{figure}
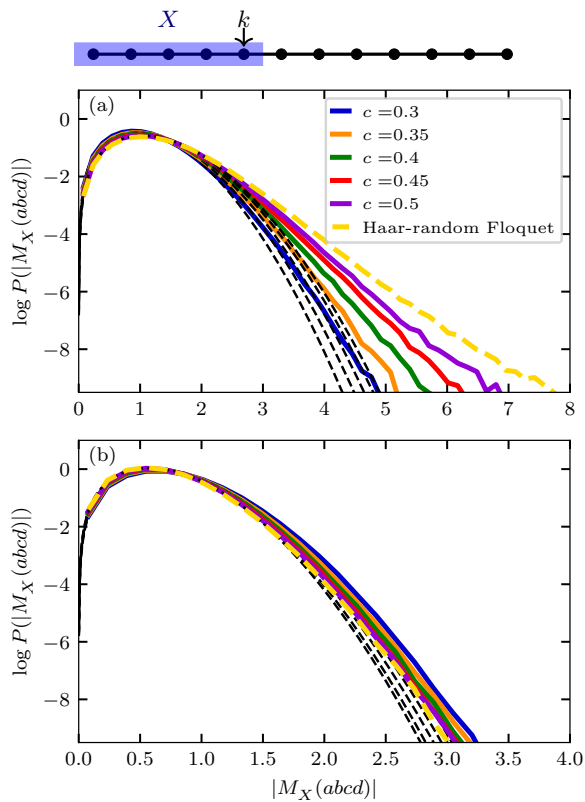
\begin{figure*}[t!]
	\centering
    %\documentclass{standalone}
%\usepackage{tikz}
%\begin{document}
        \begin{tikzpicture}
        \draw[very thick] (0.5,0) -- (6,0);
        \draw[very thick] (0.5,0.75) -- (6,0.75);
        \foreach \x in {0.5,1.0,1.5,2.0,2.5,3.0,3.5,4.0,4.5,5.0,5.5,6.0}
        \filldraw (\x,0) circle (2pt);
        \foreach \x in {0.5,1.0,1.5,2.0,2.5,3.0,3.5,4.0,4.5,5.0,5.5,6.0}
            \filldraw (\x,0.75) circle (2pt);
            \filldraw[opacity=0.4,color=blue] (0.25,0.9) rectangle (2.75,0.6);
            \filldraw[opacity=0.1,color=blue] (2.75,0.9) rectangle (6.25,0.6);
            
            \filldraw[opacity=0.1,color=red] (0.25,-0.15) rectangle (3.75,0.15);
            \filldraw[opacity=0.4,color=red] (3.75,-0.15) rectangle (6.25,0.15);
            %\draw[very thick,<->] (2.75,0.25) -- node[above] {$s$} (3.75,0.25);
            \node[anchor=north,color=blue!100!black] (x) at (2.5,0.6) {$x$};
            \node[anchor=north,color=red!100!black] (x) at (3.5,0.6) {$y$};
            \node[anchor=south,color=blue!50!black] (x) at (1.5,0.9) {$X$};
            \node[anchor=south,color=blue!50!black] (x) at (4.5,0.9) {$\overline{X}$};
            
            \node[anchor=south,color=red!50!black] (y) at (5,-0.75) {$\overline{Y}$};
            \node[anchor=south,color=red!50!black] (y) at (2,-0.75) {$Y$};
    \end{tikzpicture}
%\end{document}        
    \vspace{0.2cm}
	\includegraphics[width=0.9\textwidth]{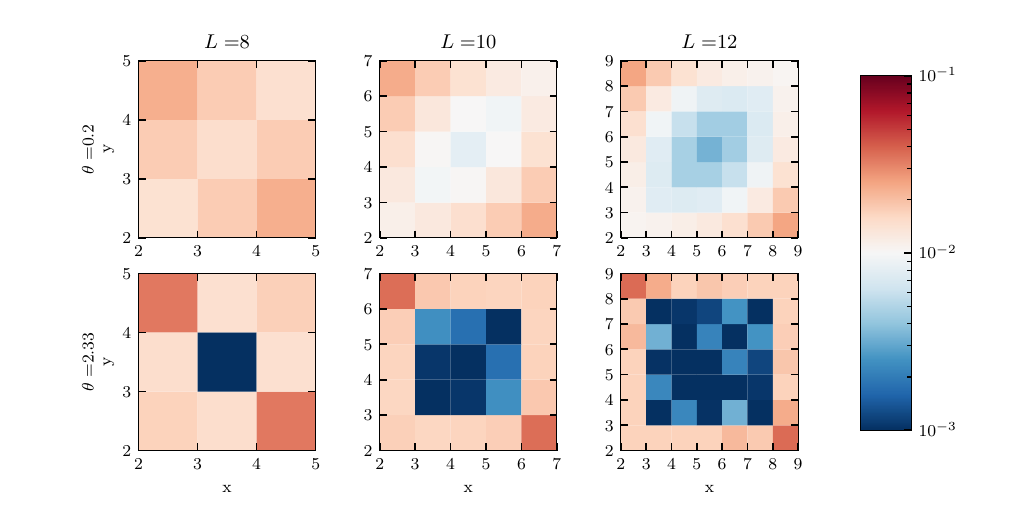}
    \caption{Relative deviation between $F^{MC}_4(X,Y,\theta)$ and $F_4(X,Y,\theta)$ for $q=2$, and $L=8$~(left column), $L=10$~(center column) and $L=12$~(right column), and for $\theta=0.1$ (top row) and $\theta=2.03$ (bottom row). 
}
\label{fig:M0rel}
\end{figure*}
In this appendix we examine the probability distribution of $M_X(abcd)$, providing further details beyond the information given in Fig.~\ref{fig:M0histo}. This information is important because one of the two methods that we use for determining the Lagrange multipliers $G_4(X,Y,\theta)$ relies on this distribution having a Gaussian form. In Appendix~\ref{subsec:accuracy} we show how the form of the distribution varies with the value of $L(X)$. In Appendix~\ref{subsec:Weak links} we investigate the effect on the distribution of the cutoff in the operator entanglement purity of two-qubit gates introduced in Sec.~\ref{subsec:Models}.

\subsection{Effect of the subsystem size $L(X)$}
\label{subsec:accuracy}

The distribution of $M_X(abcd)$ for different partitions $X$ is shown in Fig.~\ref{fig:M0histodistribution} (a) and (b). With increasing partition size $L(X)$, the distribution approaches a Gaussian. This supports our observation that the method we use to determine $G_4(X,Y,\theta$ becomes more accurate with increasing $L(X)$.

\subsection{Effect of the cutoff in the operator entanglement purity}\label{subsec:Weak links}

In Sec.~\ref{sec:Implementation} we presented results for a model (see Sec.~\ref{subsec:Models}) that is well adapted to our procedure for finding $G_4(X,Y,\theta$). This model is designed to have a probability distribution for $M_X(abcd)$ that is close to Gaussian, and has gates drawn from a truncated version of the Haar distribution, with a cutoff on the operator entanglement purity of $c\times q^4$.

In Fig.~\ref{fig:M0histodistribution2} we examine the effect of the value of this cutoff on the probability distribution of $M_X(abcd)$, considering the range from $c=0.3$ (the value used in Sec.~\ref{sec:Implementation}) to $c=1$ (an unrestricted Haar distribution). The distribution shows non-Gaussian tails for small relative phase $\theta$ and a Haar-random Floquet model. The tails are suppressed with decreasing cutoff in the operator entanglement purity.
We attribute this effect to weak links $i$ on which the gate $w_i$ is close to the identity (especially in small systems or with small subsystems). The effect of such weakly entangling gates on the dynamics of quantum information was studied in more detail in~\cite{nahum2018Dynamics}. 

\section{Accuracy of our approach with increasing system size}
\label{sec:Error scaling system size}

Here we provide complete information  on deviations between $F^{MC}_4(X,Y,\theta)$ and $F_4(X,Y,\theta)$ with Lagrange Multipliers $G_4(X,Y,\theta)$ determined using the single-shot procedure described in Sec.~\ref{G4fit}.
We compare the results for system sizes $L=8,\,10,\,12$ and 
all possible partitions defined by one cut that have $L(X),L(\overline{X})>2$.

The relative deviations are shown in Fig.~\ref{fig:M0rel}. In all cases the relative deviation is less than $10\%$, showing the accuracy of our approach. Furthermore, the accuracy improves with increasing system size. We see the largest deviations for small partition sizes $L(X)$ and $L(Y)$. This is expected, as we discuss in Sec.~\ref{sec:perturbation theory}: Correction terms in the perturbation theory and contributions from $G_2(X,\theta)$ are most significant when $L(X)$ and $L(\overline{X})$ are small.

\section{Result for single realizations}
\begin{figure}[h!]
	\centering      
	\includegraphics[width=0.45\textwidth]{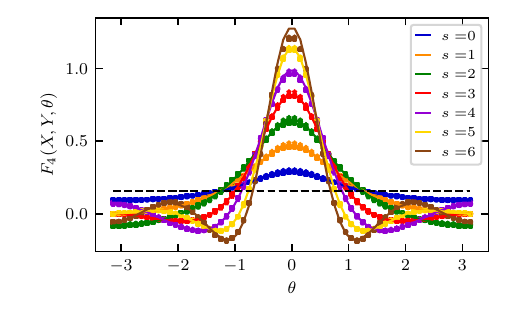}
    \caption{$F_4(X,Y,\theta)$ (lines from ED) vs $\theta$ for various $s$, and results for single Floquet realizations~(squares, stars and circles) for the circuit model with an additional cutoff in the operator purity, studied in Sec~\ref{sec:Synopsis}. The results of single Floquet realizations are close to the ensemble averages.}
	\label{fig:Singlerealization}
\end{figure}
In the main text, the eigenstate correlations obtained by exact diagonalization were averaged over multiple realizations of the Floquet circuit to suppress sample-to-sample fluctuations. This is especially important for the $q=2$ model with Haar gates studied in Sec.~\ref{sec:q2Haarresults}, where the presence of weak links can dominate the overall dynamics. 
As we show in this appendix, this is not the case for the Haar model with additional cutoff studied in Sec.~\ref{sec:Synopsis}.

In Fig.~\ref{fig:Singlerealization}, the averaged result for the eigenstate correlations $F_4(X,Y,\theta)$ together with the results for three different single realizations are shown. The deviations between different realizations of the Floquet circuits are marginal, so we expect that our analysis holds even without ensemble averaging for the circuit analyzed in Sec.~\ref{sec:Synopsis}.

\section{Results with additional Lagrange multipliers}\label{sec:doublecut}

In the main text, we have included Lagrange multipliers $G_4(X,Y,\theta)$ only for partitions $X$, $Y$  that can be generated by a single cut (apart from a discussion in Sec.~\ref{sec:q2resultspbc} of connected partitions generated by two cuts in a system with periodic boundary conditions).
The purpose of this appendix is to examine the effect of including additional partitions.
As an illustration, we include all partitions that can be generated in a system with open boundary conditions using either one or two cuts with $L(X),\,L(Y),\, L(\overline{X}),\,L(\overline{Y})>2$.
%\imk{\sout{Furthermore, we obtain $\mathsf{G}_4^{(0)}$ by using the exact Haar averages instead of the Gaussian averages in the main text. Apart from this difference, $\mathsf{G_4}$ is determined as described in Sec.~\ref{G4fit}.}}

\begin{figure}[h!]
	\centering      
	\includegraphics[width=0.45\textwidth]{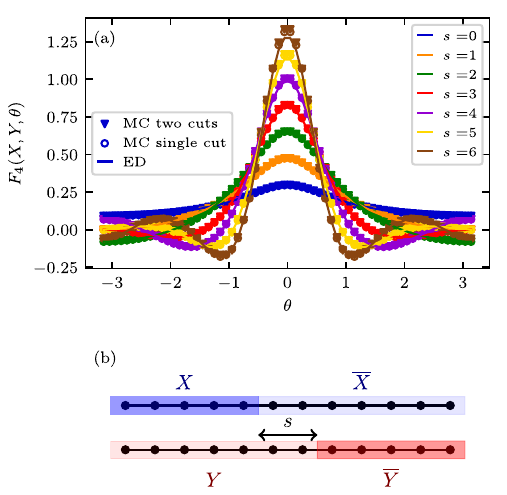}
    \caption{$F^{\rm MC}_4(X,Y,\theta)$ including all partitions generated by two cuts with $L(X)>2$,(solid triangles from MC), $F^{\rm MC}_4(X,Y,\theta)$ including all partitions generated by a single cut with $L(X)>2$ (open circles)  and $F_4(X,Y,\theta)$ (lines from ED) vs $\theta$ for various $s$. Calculations are for the brickwork circuit model defined in Sec.~\ref{subsec:Models} with $L=12$, $q=2$ and open boundary conditions.}
	\label{fig:Doublecutnormal}
\end{figure}

In Fig.~\ref{fig:Doublecutnormal} we compare the outcome in this case with results using only partitions generated by means of a single cut.
%test the results for the previously chosen partition in Fig.~\ref{fig:Doublecutnormal} and compare with the case including only a single cut. 
$F^{\rm MC}_4(X,Y,\theta)$ agrees well in both cases with
$F_4(X,Y,\theta)$ obtained from exact diagonalization. 
However, including more partitions does not reduce the small discrepancies between MC and ED results.

\begin{figure}[h!]
	\centering      
	\includegraphics[width=0.45\textwidth]{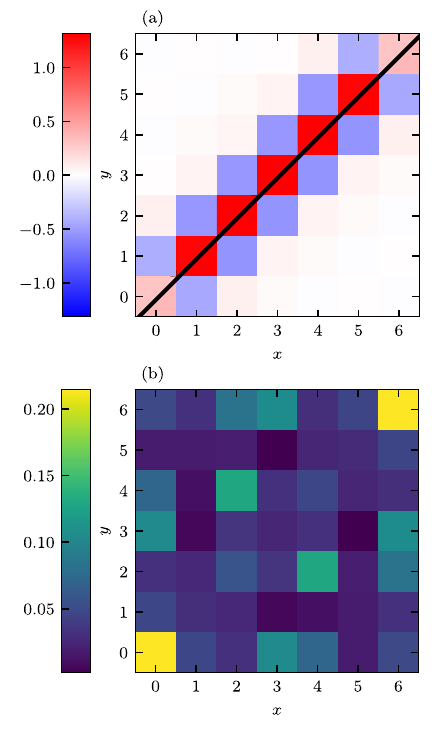}
    \caption{(a) Comparison of $S^4_{x,y}=S^4_{y,x}$ including partitions generated by a single cut~(above the diagonal) and by two cuts~(below the diagonal), for a relative phase $\theta=0.0$, $L=12$,$q=2$.
    (b) Relative deviation. Adding additional Lagrange multipliers leads to only small modifications.
	\label{fig:Gmatrix}}
\end{figure}
\begin{figure}[h!]
	\centering      
	\includegraphics[width=0.45\textwidth]{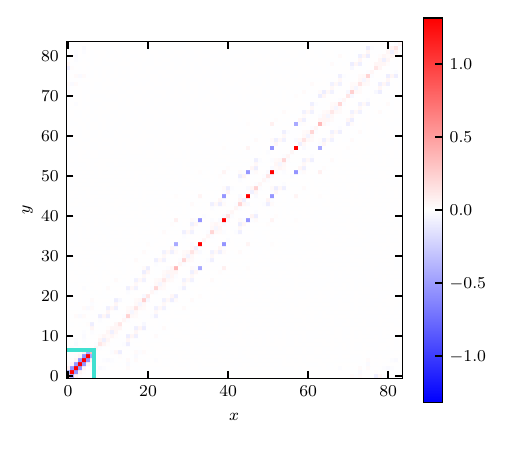}
    \caption{The complete matrix 
    %${S}^4_{x,y}$ 
    {\color{blue} ${S}_4(X,Y)$}  for the case including partitions generated by two cuts with $L(X)>2$. The submatrix analyzed in Fig.~\ref{fig:Gmatrix} is indicated as a turquoise square in the lower left corner.}
	\label{fig:Gmatrixcomplete}
\end{figure}

Finally, we investigate how the Lagrange multipliers $G_4(X,Y,\theta)$ are modified after adding additional partitions. 
To present the results it is useful to view $G_4(X,Y,\theta)$ as a matrix, as done in Sec.~\ref{G4fit}, and specify an order for the rows and columns of the matrix. We do this using the index $x=i(L-5)+(j-3)$ for a partition with support on the sites $\{i,i+1,\dots,(i+j)\mod L]\}$.
If only partitions generated by a single cut are included, $i=0$ and $j\geq3$, leading to a $(L-5)\times(L-5)$ matrix.  If on the other hand partitions generated by two cuts are included, both $i$ and $j$ may be positive, which leads to a $L(L-5)\times L(L-5)$ matrix.
Restricting to the first $L-5$ indices allows a direct comparison of the two cases.

In the following, we analyze the component-wise Haar-averaged version of Eq.~\eqref{eq:matrix version}
\begin{align}
 \mathsf{S}_4(X,Y)=\big[[\mathsf{M^T}]_X [\mathsf{G_4}]_{X,Y}[\mathsf{M^T}]_Y\big]_0.
\end{align}
As a first test, we directly compare in Fig.~\ref{fig:Gmatrix} the 
values of $\mathsf{S_4}(X,Y)$ corresponding to the first $L-5$ indices. In this case, including more partitions only leads to small modifications of these Lagrange multipliers.

Going further, in Fig.~\ref{fig:Gmatrixcomplete} we examine the extra contributions to $\mathsf{S_4}(X,Y)$ that arise after incorporating partitions generated by two cuts. In this figure the Lagrange multipliers related to single-cut generated partitions contribute to the small turquoise square at the lower left corner, and so we are concerned with the remainder of the figure. As is apparent, the values of $\mathsf{S_4}(X,Y)$ arising from most additional partitions are very small in comparison to those from single-cut partitions. A few exceptions are also visible: they appear to arise from partitions in which there are two scales for spatial separations, such as (in obvious notation) the partitions $XXXIIIIIIXXX$ and $IIIIYYYYIIII$.
\bibliography{references,refs_eth_corr}

\end{document}